\documentclass[twocolumn,twocolappendix,iop,apj]{aastex631}
\usepackage{color,graphicx,natbib}
\usepackage{graphicx}
\usepackage{amsmath}
\usepackage{amssymb}
\usepackage{bm}
\usepackage[T1]{fontenc}
\usepackage{hyperref}
\hypersetup{
    colorlinks=True,
    citecolor=black,
    linkcolor=black,
    }

\newcommand{\Swift}{\textit{Swift}}
\newcommand{\Konus}{\textit{Konus-Wind}}
\newcommand{\Fermi}{\textit{Fermi}}

\newcommand{\Chandra}{\textit{Chandra}}
\newcommand{\XMM}{\textit{XMM-Newton}}
\newcommand{\INTEGRAL}{\textit{INTEGRAL}}
\newcommand{\HST}{\textit{HST}}
\newcommand{\EK}{\ensuremath{E_{\rm K}}}
\newcommand{\EKiso}{\ensuremath{E_{\rm K,iso}}}

\newcommand{\Egammaiso}{\ensuremath{E_{\gamma,\rm iso}}}	     

\newcommand{\epse}{\ensuremath{\epsilon_{\rm e}}}
\newcommand{\epsb}{\ensuremath{\epsilon_{\rm B}}}
\newcommand{\dens}{\ensuremath{n_{0}}}

\newcommand{\tjet}{\ensuremath{t_{\rm jet}}}
\newcommand{\thetajet}{\ensuremath{\theta_{\rm jet}}}

\newcommand{\AV}{\ensuremath{A_{\rm V}}}

\newcommand{\pcc}{\ensuremath{{\rm cm}^{-3}}}

\newcommand{\p}{\ensuremath{^{\prime}}}

\newcommand{\nua}{\ensuremath{\nu_{\rm a}}}

\newcommand{\numax}{\ensuremath{\nu_{\rm m}}}
\newcommand{\nuc}{\ensuremath{\nu_{\rm c}}}

\newcommand{\nux}{\ensuremath{\nu_{\rm X}}}

\newcommand{\erg}{~\ensuremath{\rm erg}}
\newcommand{\cm}{~\ensuremath{\rm cm}}
\newcommand{\s}{~\ensuremath{\rm s}}

\newcommand{\fnumax}{\ensuremath{F_{\nu,\rm m}}}

\newcommand{\grb}{GRB~211106A}

\newcommand{\nuchat}{\ensuremath{\hat{\nu}_{\rm c}}}
\shorttitle{GRB~211106A}
\shortauthors{Laskar et al.}

\begin{document}
\title{The First Short GRB Millimeter Afterglow: The Wide-Angled Jet of the Extremely Energetic SGRB~211106A}

\author[0000-0003-1792-2338]{Tanmoy Laskar}
\affil{Department of Astrophysics/IMAPP, Radboud University, PO Box 9010,
6500 GL, The Netherlands}
\affil{Department of Physics \& Astronomy, University of Utah, Salt Lake City, UT 84112, USA}

\author[0000-0003-3937-0618]{Alicia Rouco Escorial}
\affil{Center for Interdisciplinary Exploration and Research in Astrophysics (CIERA) and Department of Physics and Astronomy, Northwestern University, Evanston, IL 60208, USA}

\author[0000-0001-9915-8147]{Genevieve Schroeder}
\affil{Center for Interdisciplinary Exploration and Research in Astrophysics (CIERA) and Department of Physics and Astronomy, Northwestern University, Evanston, IL 60208, USA}

\author[0000-0002-7374-935X]{Wen-fai Fong}
\affil{Center for Interdisciplinary Exploration and Research in Astrophysics (CIERA) and Department of Physics and Astronomy, Northwestern University, Evanston, IL 60208, USA}

\author[0000-0002-9392-9681]{Edo Berger}
\affil{Center for Astrophysics | Harvard \& Smithsonian, 60 Garden St. Cambridge, MA 02138, USA}

\author[0000-0002-2149-9846]{P\'eter Veres} 
\affil{Center for Space Plasma and Aeronomic Research, University of Alabama in Huntsville, 320 Sparkman Drive, Huntsville, AL 35899,}

\author[0000-0003-3460-506X]{Shivani Bhandari}
\affil{ASTRON, Netherlands Institute for Radio Astronomy, Oude Hoogeveensedijk 4, 7991 PD
Dwingeloo, The Netherlands}
\affil{Joint institute for VLBI ERIC, 
Oude Hoogeveensedijk 4, 7991 PD Dwingeloo, The Netherlands}
\affil{Anton Pannekoek Institute for Astronomy, University of Amsterdam, Science Park 904, 1098 XH, Amsterdam, The Netherlands}
\affil{CSIRO Space and Astronomy, Australia Telescope National Facility, PO Box 76, Epping, NSW 1710, Australia}

\author[0000-0002-9267-6213]{Jillian Rastinejad}
\affil{Center for Interdisciplinary Exploration and Research in Astrophysics (CIERA) and Department of Physics and Astronomy, Northwestern University, Evanston, IL 60208, USA}

\author[0000-0002-5740-7747]{Charles D. Kilpatrick}
\affil{Center for Interdisciplinary Exploration and Research in Astrophysics (CIERA) and Department of Physics and Astronomy, Northwestern University, Evanston, IL 60208, USA}

\author[0000-0002-2810-8764]{Aaron Tohuvavohu}
\affil{David A. Dunlap Department of Astronomy \& Astrophysics, University of Toronto, Toronto, ON, Canada}

\author[0000-0003-4768-7586]{Raffaella Margutti}
\affil{Department of Astronomy, University of California, 501 Campbell Hall, Berkeley, CA 94720-3411, USA}


\author[0000-0002-8297-2473]{Kate D. Alexander}
\affil{Center for Interdisciplinary Exploration and Research in Astrophysics (CIERA) and Department of Physics and Astronomy, Northwestern University, Evanston, IL 60208, USA}

\author[0000-0001-5229-1995]{James DeLaunay}
\affil{Department of Physics, Pennsylvania State University, University Park, PA 16802, USA}
\affil{Center for Multimessenger Astrophysics, Institute for Gravitation and the Cosmos, Pennsylvania State University, University Park, PA 16802, USA}
\affil{Department of Physics \& Astronomy, University of Alabama, Tuscaloosa, AL 35487, USA}

\author[0000-0002-6745-4790]{Jamie A. Kennea}
\affil{Department of Astronomy and Astrophysics, The Pennsylvania State University, 525 Davey Lab, University Park, PA 16802, USA}

\author[0000-0002-2028-9329]{Anya Nugent}
\affil{Center for Interdisciplinary Exploration and Research in Astrophysics (CIERA) and Department of Physics and Astronomy, Northwestern University, Evanston, IL 60208, USA}

\author[0000-0001-8340-3486]{K. Paterson}
\affil{Max-Planck-Institut f\"ur Astronomie (MPIA), Königstuhl 17, 69117 Heidelberg, Germany}

\author[0000-0003-3734-3587]{Peter K.\ G.\ Williams}
\affil{Center for Astrophysics | Harvard \& Smithsonian, 60 Garden St. Cambridge, MA 02138, USA}
\affil{American Astronomical Society, 1667 K St. NW Ste. 800, Washington, DC 20006}

\begin{abstract}
We present the discovery of the first millimeter afterglow of a short-duration $\gamma$-ray burst (SGRB) and the first confirmed afterglow of an SGRB localized by the GUANO system on \Swift. Our Atacama Large Millimeter/Sub-millimeter Array (ALMA) detection of SGRB~211106A establishes an origin in a faint host galaxy detected in Hubble Space Telescope (\HST) imaging at $0.7\lesssim z\lesssim1.4$.
From the lack of a detectable optical afterglow, coupled with the bright millimeter counterpart, we infer a high extinction, $\AV\gtrsim2.6$~mag along the line of sight, making this the one of the most highly dust-extincted SGRBs known to date.
The millimeter-band light curve captures the passage of the synchrotron peak from the afterglow forward shock and reveals a jet break at $\tjet=29.2^{+4.5}_{-4.0}$~days.
For a presumed redshift of $z=1$, we infer an opening angle, $\thetajet=(15.5\pm1.4)$~degrees, and beaming-corrected kinetic energy of $\log(\EK/\erg)=51.8\pm0.3$, making this one of the widest and most energetic SGRB jets known to date. Combining all published millimeter-band upper limits in conjunction with the energetics for a large sample of SGRBs, we find that energetic outflows in high density environments are more likely to have detectable millimeter counterparts. Concerted afterglow searches with ALMA should yield detection fractions of 24--40\% on timescales of $\gtrsim2$~days at rates $\approx0.8$--1.6 per year,  outpacing the historical discovery rate of SGRB centimeter-band afterglows. 
\end{abstract}

\section{Introduction}
\label{text:intro}

Short-duration $\gamma$-ray bursts (SGRBs) are produced in the mergers of compact objects involving a neutron star \citep{ber14,aaa+17b}. These explosive transient events are a known site of $r$-process nucleosynthesis and thus a source of heavy elements \citep{bfc13,tlf+13,kmb+17}. Their association with gravitational wave transients makes excellent probes of fundamental physics, from cosmology to Lorentz violation \citep{mc21}. 

The interaction of the collimated, relativistic jets of SGRBs with the environment produces relativistic shocks, which accelerate electrons and produce the synchrotron afterglow \citep{gps99,gs02}. In addition to providing precise localizations (and hence, host associations and redshifts), observing and modeling afterglow emission yields the explosion energy, density and density profile of the pre-explosion environment, and the degree of ejecta collimation (e.g.~\citealt{fbmz15}). These measurements enable tests of progenitor models, delay time distributions, and true event rates corrected for beaming \citep{ber14}. 

Afterglow observations at millimeter (mm) wavelengths probe the synchrotron peak, which is sensitive to the explosion energy and density. Furthermore, mm-band observations are  unaffected by interstellar scintillation, thermal emission from the supernova/kilonova/host galaxy, and inverse Compton effects, which can impact observations at centimeter (cm), optical, and X-ray bands, respectively. 
In contrast to cm wavelengths, the mm-band is also not subject to synchrotron self-absorption at the low densities ($\lesssim 1$\pcc) typical of SGRB environments, making it an excellent wavelength to probe the location and evolution of the peak of the spectral energy distribution (SED). 

Observations of the mm afterglows of long-duration $\gamma$-ray bursts (LGRBs, originating in the deaths of massive stars; \citealt{wb06}) with the improved sensitivity of the Atacama Large Millimeter/Submillimeter Array (ALMA) are already proving revolutionary, and have resulted in (i) confident detection and characterization of reverse shocks \citep{lab+16,lves+19}; (ii) constraints on the degree of GRB
ejecta magnetization \citep{lag+19}; and (iii) studies of ejecta collimation \citep{lab+18}. However, no mm-band afterglow for an SGRB has been reported to date. The deepest limits from prior to the commissioning of ALMA are comparable to the observed mm-band luminosity of the least luminous LGRB afterglows \citep{cdupg+05, duplm+12}. 
While deeper mm-band limits have been published for GRB~170817A \citep{abf2017,ksr+17}, these limits were not constraining, as the peak of the SED was already below the cm-band at the time.
The lower energy and lower ambient density of SGRBs compared to LGRBs are expected to reduce the peak fluxes of their mm afterglows, putting them largely out of reach of these pre-ALMA facilities \citep{duplm+12, phct+19}. 

Here, we present the discovery of the mm afterglow of \grb. We discuss our $\gamma$-ray to radio observations of this burst in  Section~\ref{text:observations}. We associate the burst with a host galaxy and consider its properties in Section~\ref{text:host}. We perform multi-wavelength afterglow modeling in Section~\ref{text:modeling} and discuss the results in Section~\ref{text:discussion}. 
No redshift is available for this event, and, where relevant, we perform our analysis at two fiducial redshift values of $z=0.5$ and $z=1$, which are chosen to approximately correspond to the median values of large spectroscopic and photometric samples of SGRB hosts, respectively \citep{fnd+22,nfd+22}, and report the results from both. We use a $\Lambda$CDM cosmology with $\Omega_{\rm M}=0.31$, $\Omega_{\Lambda}=0.69$, and $h=0.68$ throughout. All magnitudes reported here are in the AB system and not corrected for Galactic extinction. All uncertainties are $1\sigma$ and upper limits are $3\sigma$, unless otherwise noted. 

\begin{figure*}
\includegraphics[width=0.9\columnwidth]{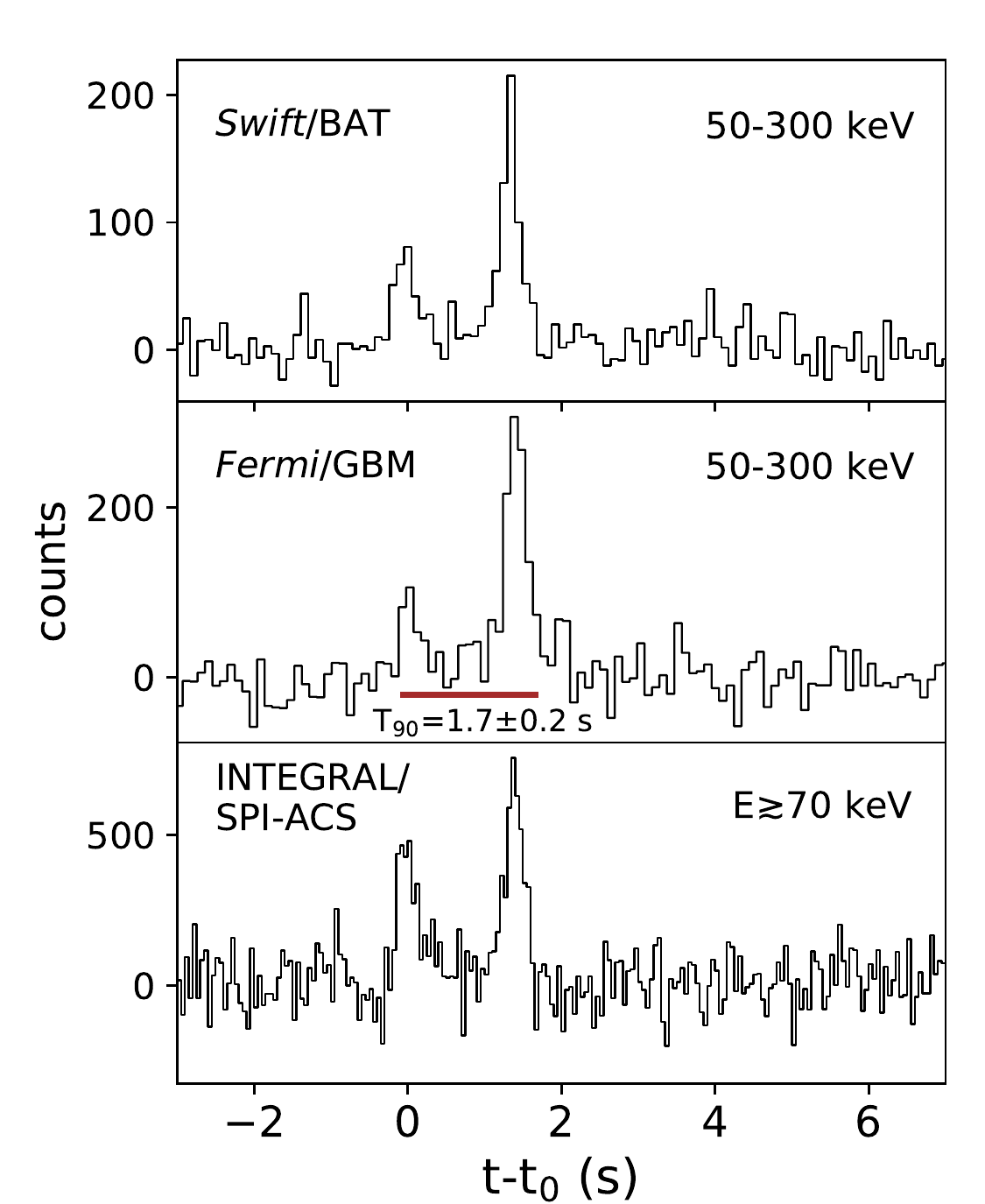}
\includegraphics[width=1.1\columnwidth]{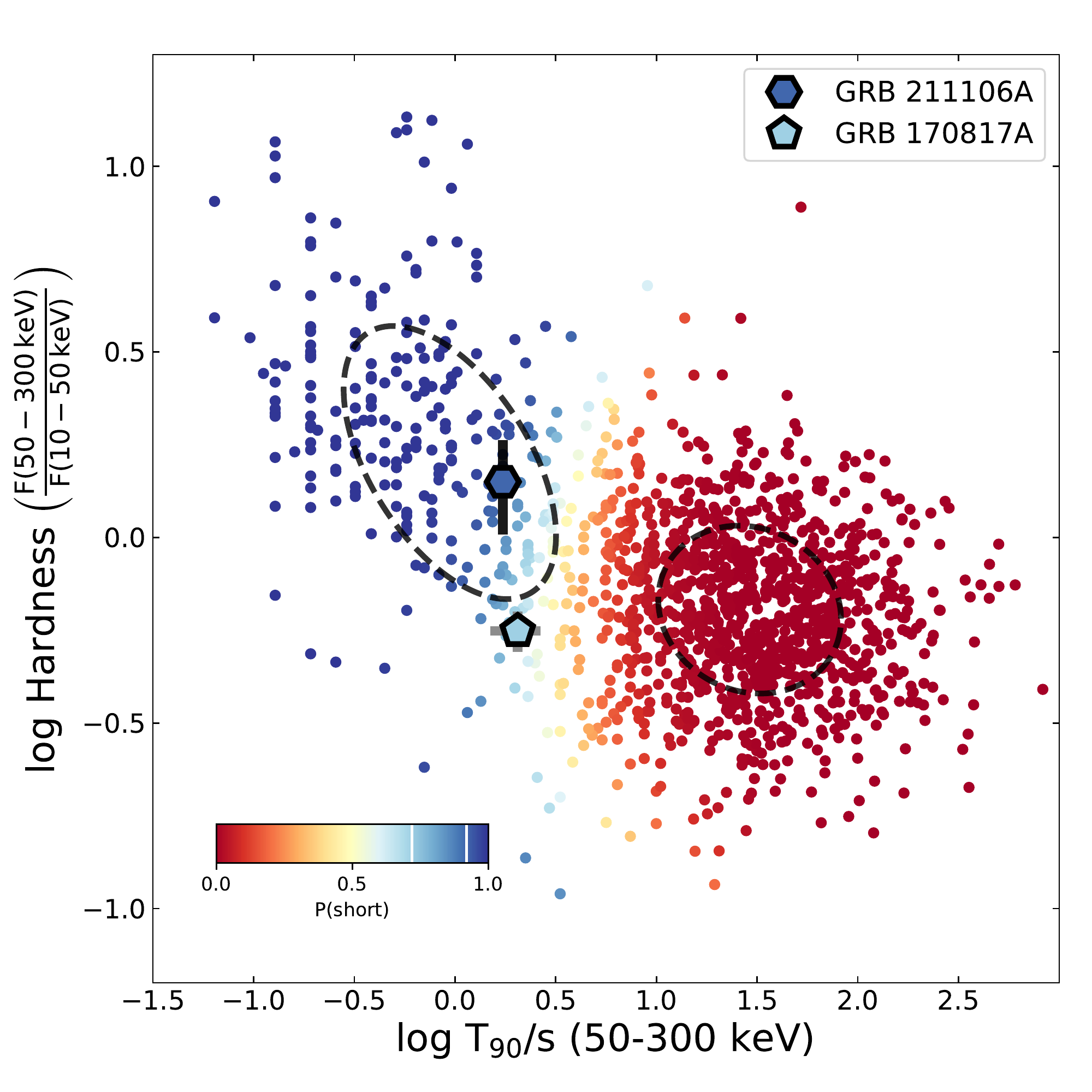}
\caption{{\it Left:} background-subtracted $\gamma$-ray light curves of \grb\ in a common reference frame corrected for light-travel time effects in the canonical 50-300~keV range from \Swift/BAT (top) and \Fermi/GBM (middle; both at 96~ms resolution) and at $E\gtrsim70$ keV from INTEGRAL/SPI-ACS (bottom) with 50~ms resolution. 
{\it Right:} The location of this event in the duration-hardness plane of Fermi GRBs \citep{bmv16} colored by $P_{\rm short}$ indicates \grb\ has a high ($\approx92\%$) likelihood of belonging to the SGRB population (Section~\ref{text:gamma-ray}). White lines in the color bar refer to the values of $P_{\rm short}\approx0.92$ and $P_{\rm short}\approx 0.72$ for \grb\ and GRB~170817A, respectively (Section~\ref{text:gamma-ray}).}
\label{fig:gamma-ray}
\end{figure*}

\section{Observations and Data Analysis}
\label{text:observations}

\subsection{$\gamma$-ray: \Swift/BAT and \Fermi/GBM analysis}
\label{text:gamma-ray}
\label{text:discovery}
Prompt $\gamma$-rays from \grb\ were first discovered\footnote{\url{https://gcn.gsfc.nasa.gov/other/9504.integral_spiacs}} by \INTEGRAL/SPI-ACS \citep{vbr+03} on 2021 November 6 at 04:37:31.2 UT \citep{gcn31049}; all times in this paper are given relative to this time. 
The \INTEGRAL\ notice triggered the Gamma-ray Urgent Archiver for Novel Opportunities \citep[GUANO,][]{tkd+20} operated by the Neil Gehrels Swift Observatory's (\Swift; \citealt{gcg+04}) Mission Operations Center. GUANO ordered the \Swift\ Burst Alert Telescope \citep[BAT,][]{bbc+05} to save 90\,s of BAT event-mode data around the time of burst. Using the Non-Imaging Transient Reconstruction And TEmporal Search \citep[NITRATES; ][]{dt21}, we find a strong detection with BAT in the time-domain. The BAT light curve exhibits two pulses, with duration, $T_{90}=1.7\pm0.1$\,s ($50-300$\,keV, observer frame; Fig.~\ref{fig:gamma-ray}). 
The best-fit BAT position\footnote{See Appendix~\ref{text:nitrates} for details of the NITRATES localization.} from NITRATES is RA = 22h\,54m\,34.32s and Dec = $-53$d\,14\arcmin\,0.9\arcsec, with an uncertainty of 7\arcmin\ \citep{gcn31049}.

\grb\ also triggered \Konus\ \citep{afg+95} on 2021 November 6 at 04:37:32.485 UT. The observation revealed a light curve\footnote{\url{http://www.ioffe.ru/LEA/GRBs/GRB211106_T16652/}} with a single-pulse structure of $\sim0.5$\,s (20\,keV$-$2\,MeV), consistent with the time of the second peak in the BAT light curve \citep{gcn31054}. 

The Gamma-ray Burst Monitor \citep[GBM,][]{mlb+09} on-board \Fermi\ \citep{GLAST1999}, did not trigger on this GRB. \citet{ff21} identified a significant event (signal-to-noise of 22) in the GBM data using the off-line {\tt targeted search} pipeline \citep{ghw+19},  
at a position consistent with the \Swift/BAT-GUANO position.
The {\it Fermi}-GBM light curve exhibits two pulses coincident with those in the BAT light curve, and with $T_{90}=1.71\pm 0.18 \s$ (50--300 keV, observer frame; Fig.~\ref{fig:gamma-ray}). Fitting the time-integrated GBM spectrum during the $T_{90}$ interval with {\tt RMfit}\footnote{\url{ https://fermi.gsfc.nasa.gov/ssc/data/analysis/rmfit}} using a power law model with an exponential cutoff (parameterized as a peak energy is $E_{\rm peak}$), we find a photon index, $\Gamma_{\rm \gamma,\rm CPL}=-0.85\pm 0.20$, $E_{\rm peak}=306\pm60$~keV, and $\gamma$-ray fluence,  $\mathcal{F_\gamma}=(1.56\pm0.14)\times 10^{-6} \erg \cm^{-2}$. 
The isotropic-equivalent $\gamma$-ray energy (1--$10^4$ keV, rest frame) corresponds to $E_{\gamma,\rm iso}=(1.1\pm0.1)\times 10^{51} \erg$ at $z=0.5$ and $E_{\gamma,\rm iso}=(4.4\pm0.4)\times 10^{51} \erg$ at $z=1.0$, two redshifts spanning the typical range for SGRBs as discussed in Section \ref{text:intro}. 
Both \Egammaiso\ estimates are consistent with the $E_{\rm peak}$--$E_{\gamma,\rm iso}$ distributions for SGRBs \citep{gng+09,tyn+13,mp20}. 

We compute the spectral lag between low-energy (25--50~keV) and high-energy (100 -- 300 keV) GBM light curves 
using the same energy bands and procedure as described in \citet{nmb00}, and find 
$\tau=-35.7^{+56.1}_{-58.9}~{\rm ms}$. This is consistent with $\tau\approx0$ as measured for SGRBs \citep{gnb+06}. According to the lag-luminosity relationship for long GRBs, $L_{\rm peak}\propto(\tau/(1+z))^{-0.74}$ \citep{nmb00}, where $L_{\rm peak}$ is the peak luminosity. Fitting the brightest 0.128~s time-bin, we find a peak flux, $F_{\rm peak}\approx2.4\times 10^{-6} \erg \cm^{-2} \s^{-1}$, yielding $L_{\rm peak}\approx1.5\times 10^{51}\erg \s^{-1}$ and $6.2\times 10^{51}\erg \s^{-1}$ at $z=0.5$ and $z=1.0$ respectively. If \grb\ were a long GRB, for the measured $L_{\rm peak}$ the lag-luminosity relation would imply a lag $\tau\approx 0.238\s$ in the more conservative $z=1.0$ case. This is inconsistent with the measured lag at a $4.9~\sigma$ level.
We derive the hardness ratio (HR), defined as the photon flux above background in a high-energy band 
divided by those in a low energy band \citep{bmv16,gvb+17}, 
and find ${\rm HR}=1.41\pm 0.36$. Modeling the $T_{90}$-HR plane with a Gaussian mixture model\footnote{Details of the Gaussian mixture model are presented in Appendix \ref{text:prob}.} following \cite{rfv+21}, we find that the probability that \grb\ belongs to the SGRB population is $P({\rm short})\approx 92\%$ (Fig.~\ref{fig:gamma-ray}). 

Owing to its short $T_{90}$, hard spectrum, and negligible spectral lag, we consider \grb\ to be a bona-fide short-duration, spectrally-hard GRB.

\subsection{X-ray: \Swift, \Chandra, and \XMM}
\label{text:x-ray}

\Swift/XRT began follow-up observations of the BAT/NITRATES position at $\approx 0.46$ days, revealing a fading X-ray afterglow at RA = 22h\,54m\,20.45s and Dec = $-53$d\,13\arcmin\,49.0\arcsec, with an uncertainty of 3.4\arcsec\ \citep[90\% confidence;][]{gcn31068}. We downloaded time-sliced X-ray spectra per bin of the dynamically binned XRT light curve with the spectral extraction tool\footnote{\url{https://www.swift.ac.uk/xrt_spectra/00021466/}} on the \Swift\ website \citep{ebp+09}, which we later use together with all available X-ray data for a joint spectral analysis.

We observed the afterglow with \Chandra/ACIS-S3 \citep{gbf+03}
at $\approx 10.5$ 
and $59.8$ days 
with total effective exposure times of $19.8$\,ks and $37.9$\,ks, respectively, through target of opportunity and DDT programs \#22500107 (PI: Berger, ObsID~23543) and \#22408828 (PI: Rouco Escorial, ObsIDs~26249 and 26262). We used the \texttt{CIAO} software package \citep[v.\,4.12,][]{fma+06} and calibration files (\texttt{caldb}; v.\,4.9.0) to reduce the data.
We detect the X-ray afterglow in the first \Chandra\ epoch at RA = 22h\,54m\,20.51s and Dec = $-53$d\,13\arcmin\,51.17\arcsec\ ($1\sigma$ uncertainty of 0.62\arcsec; including centroiding uncertainty of 0.18\arcsec and absolute astrometric uncertainty of 0.6\arcsec).
We refine this position by astrometric calibration against Gaia using \HST\ imaging (Section~\ref{text:hst}) in  Appendix~\ref{text:astrometry}. 
We derive the X-ray count rate and spectrum from a 2\arcsec\ aperture centered on the X-ray afterglow and report the results in Table\,\ref{tab:data:xray}.

We used \XMM/EPIC \citep{sbd+01,taa+01} to obtain two epochs of the afterglow 
at mid-times of $\approx14.9$ and 33.0~days 
after the burst, with total effective exposure times of $20.3$\,ks and $46.7$\,ks, respectively, through target of opportunity Program \#086286 (PI: Fong, ObsIDs: 0862860301 and 0862860401). We reduced and analyzed the \XMM\ data using \texttt{SAS} \citep[v.\,18.0.0;][]{gdfh+04}. The afterglow was detected in both epochs. We derive the source flux  and spectrum using a 20\arcsec\ aperture (Table\,\ref{tab:data:xray}).

\begin{figure}
 \centering
 \includegraphics[width=1.01\columnwidth]{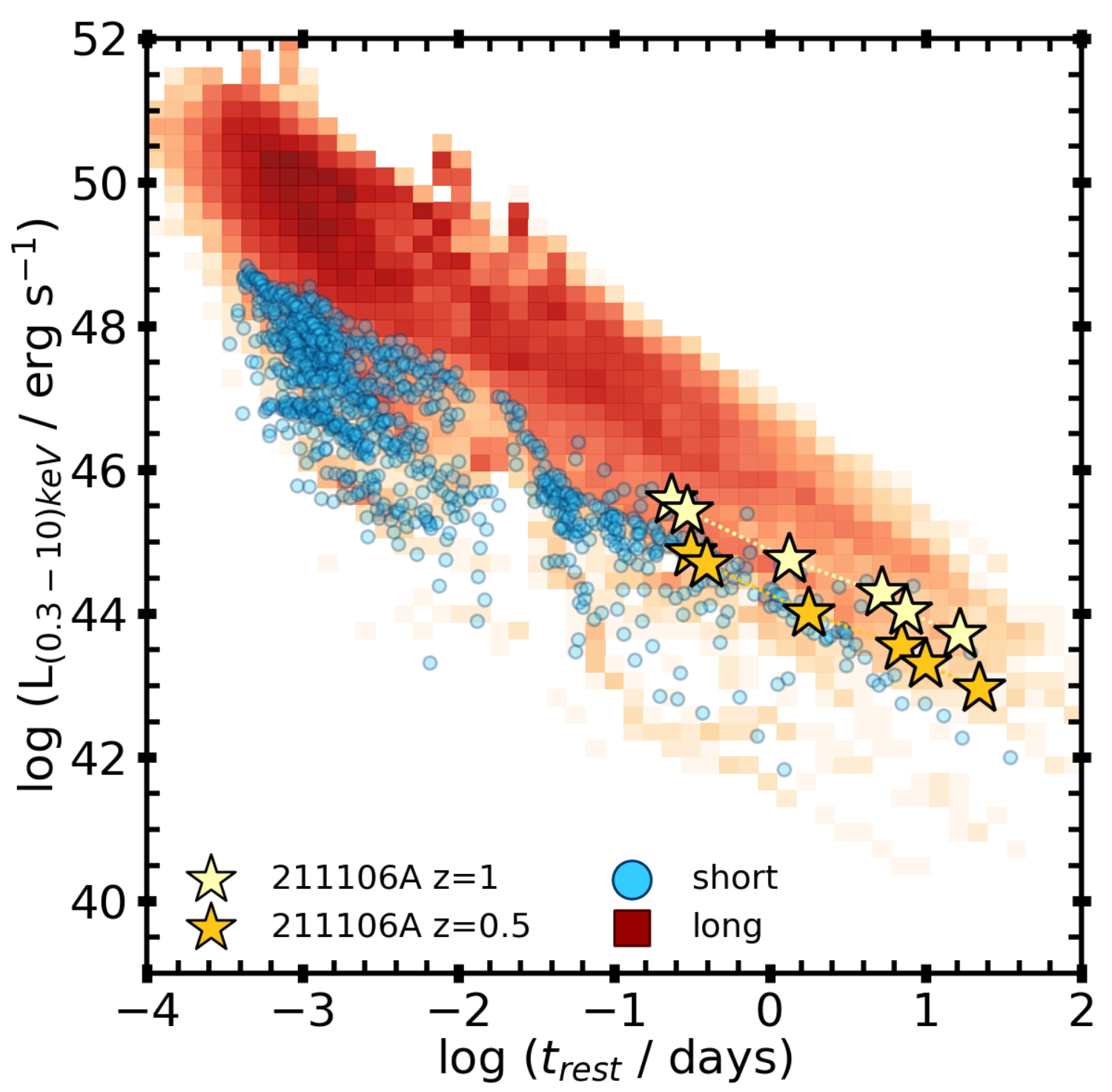}
\vspace{-0.1in}
\caption{The X-ray luminosity ($0.3-10.0$\,keV; unabsorbed (i.e., corrected for galactic and intrinsinc absorption), observer frame) versus rest-frame time for \grb\ at $z=0.5$ and $z=1$, compared with that of \Swift/BAT LGRBs (red density) and SGRBs (blue circles) with known redshifts. \grb\ exhibits one of the most luminous X-ray afterglows of the SGRB population to date.}
\label{fig:Xraylc}
\end{figure}

We use \texttt{Xspec} \citep[v.~12.10.1f;][]{a96} to perform a joint spectral fit of the \Swift, \Chandra, and \XMM\ data in the $0.5-7$~keV energy range sampled by all instruments. We use an absorbed power-law model with photon index ($\Gamma_X$), intrinsic absorption ($N_{\rm H, int}$), fixed Galactic absorption ($N_{\rm H, Gal}=1.06\times10^{20}$\,cm$^{-2}$; \citealt{wsb+13}), fixed normalization factors\footnote{Following Table~5 in \citet{pbf+17} and relative to \Chandra/ACIS-S3, these constants are 0.87, 0.90, 0.98, and 1.0 for XRT-PC, EPIC-pn, MOS1, and MOS2, respectively.} to account for cross-calibration between observatories,  W-statistics (\texttt{statistic cstat}; \citealt{wlk79}) and \texttt{WILM} abundances \citep{wam00}. 
We find no evidence for spectral evolution, and derive $\Gamma_{\rm X}=1.9\pm0.3$, 
$N_{\rm H, int}=\left(6.3^{+3.7}_{-3.2}\right)\times10^{21}~\pcc$
at $z=0.5$ and 
$N_{\rm H, int}=\left(13^{+8}_{-7}\right)\times10^{21}~\pcc$ at $z=1$. We derive unabsorbed X-ray fluxes ($0.3-10$\,keV) using the \texttt{cflux} convolution model and convert count rate upper limits to flux limits using the associated instrumental response files and Poisson statistics with the spectral parameters fixed to the best-fit values. This appears to be one of the most luminous SGRB afterglows at the corresponding rest-frame time known to date (Fig.~\ref{fig:Xraylc}). We discuss this X-ray light curve in the context of those from other SGRB in Section~\ref{text:discussion}.

\begin{figure*}
 \includegraphics[width=\textwidth]{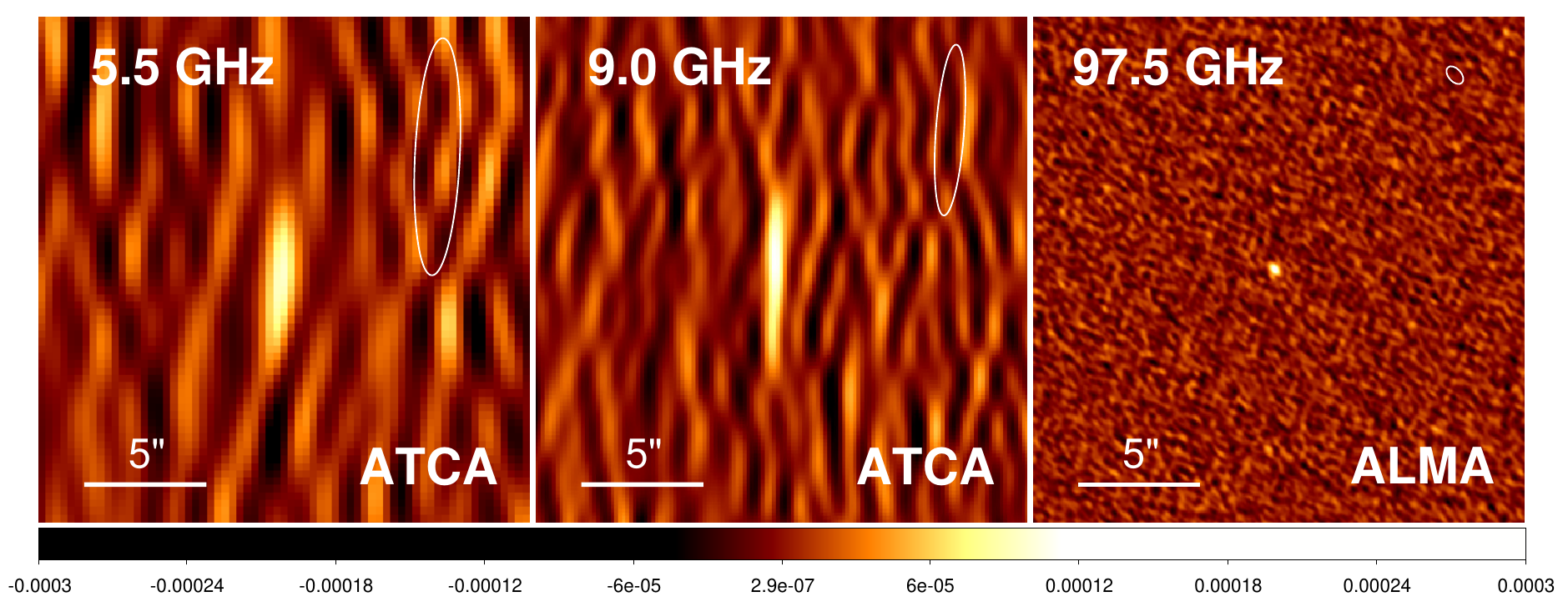}
 \caption{Discovery images of the radio and mm-band afterglow of GRB~211106A with ATCA at 5.5~GHz (left) and 9.0~GHz (center) at $\approx14.18$~days, and with ALMA at 97.5~GHz at $\approx12.89$~days after the burst. Ellipses in the top right represent the synthesized beam. The radio afterglow is clearly detected in each image. All images have the same display stretch and color scale, indicated by the color bar (in Jy) at the bottom.}
\label{fig:radioimages}
\end{figure*}

\subsection{Millimeter: ALMA}
\begin{figure*}
\centering
\includegraphics[width=\textwidth]{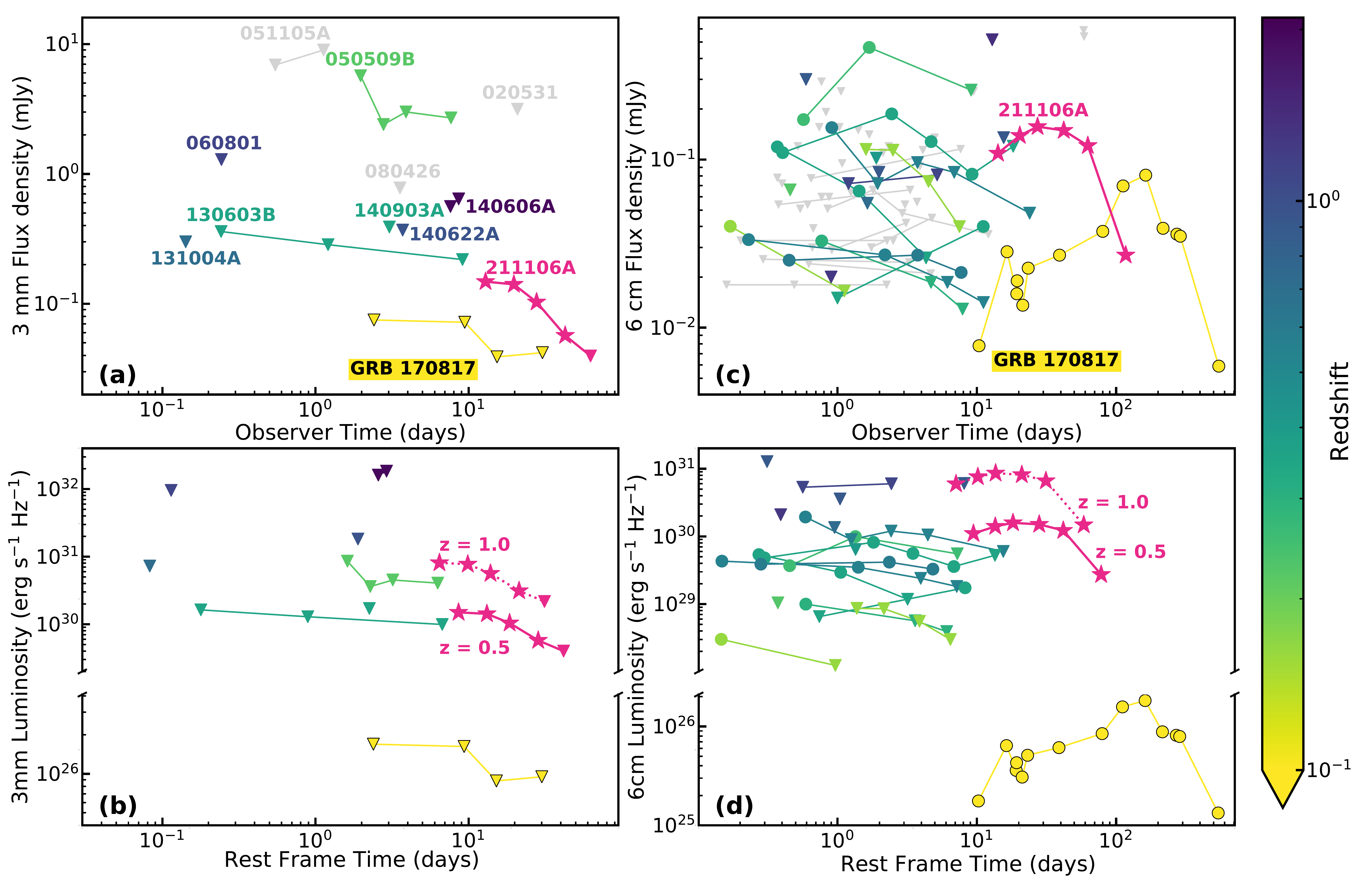}
\caption{{\it (a)}: Comparison of the mm-band afterglow of \grb\ (magenta stars) with all published 3mm light curves of SGRBs colored by redshift. Events with no known redshift are in grey. Triangles denote $3\sigma$ upper limits. After GRB~170817A, our ALMA observations of \grb\ are the deepest obtained for any SGRB to-date. 
{\it (b)}: Millimeter-band luminosity vs rest-frame time for GRB~211106A at two assumed redshifts, $z = 0.5$ (solid line) and $z = 1.0$ (dotted line), compared to SGRBs with available redshifts. 
Colors are the same as the above plot.
{\it (c)}: The 6 cm (5.5~GHz) ATCA light curve of the afterglow of GRB~211106A (magenta stars). For comparison, we show the cm-band (5-10~GHz) light curves of the 9 other radio detected SGRBs as well as GRB~170817A (circles) colored by host galaxy redshift from yellow (low) to purple (high). Triangles denote $3\sigma$ upper limits, and SGRBs with no known redshift are in grey.
{\it (d)}: Radio luminosity of the 6~cm (5.5~GHz) ATCA light curve of the afterglow of GRB~211106A vs. rest-frame time for two redshifts: $z = 0.5$ (solid line) and $z = 1.0$ (dotted line). For comparison, we show the cm-band (5-10~GHz) radio luminosity of the 9 other radio detected SGRBs as well as GRB~170817A (circles). Colors are the same as the above plot.}
\label{fig:mm_cm_comparison}
\end{figure*}

We observed \grb\ with the Atacama Large Millimeter/Submillimeter Array (ALMA) at 97.5 GHz 
at a mid-time of 12.9~days after the burst (project 2019.1.00863.T, PI: Fong). We utilized four 2~GHz spectral windows centered at $90.52$, $92.42$, $102.52$, and $104.48$ GHz, and employed J2357-5311 as bandpass and flux density calibrator, J2239-5701 as complex gain calibrator, and J2207-5346 as a check source.  We calibrated the data using the automated ALMA pipeline  \texttt{procedure\_hifa\_cal.xml} in the Common Astronomy Software Applications (CASA; \citealt{mws+07}) v.~5.6.1-8 followed by imaging to the half-power point of the primary beam using one Taylor term and with Briggs weighting using a robust parameter of 0.5. We detect a single point source with flux density $148\pm11\,\mu$Jy in the image spanning 1.5\arcmin\ in diameter (Fig.~\ref{fig:radioimages}). We obtained 4 additional epochs of ALMA observations and the mm-band point source is observed to fade to a flux density below detection by the time of the final epoch obtained 62.6~days post-burst.
The most precise position of the counterpart is afforded by the second epoch, which has the smallest synthesized beam area of $0.346\arcsec\times0.269\arcsec$, 
RA = 22h\,54m\,20.53056s ($\pm0.0012$~s, 0.011\arcsec), 
Dec = $-53$d\,13\arcmin\,50.525\arcsec\ ($\pm 0.010\arcsec)$. The absolute systematic astrometric uncertainty on this position is given by ${\rm beam}_{\rm FWHM} / SNR / 0.9 \approx36$~mas \citep{rbc+19}, with negligible additional systematic uncertainty ($\lesssim2$~mas) from the calibrator positions. The mm-band position is consistent with both the original and refined \Chandra\ afterglow position (Section~\ref{text:x-ray} and Appendix~\ref{text:astrometry}). The positional coincidence and fading behavior confirm this source as the mm afterglow of \grb. 
We plot the ALMA light curve in Fig.~\ref{fig:mm_cm_comparison} and report the corresponding flux density values in Table~\ref{tab:data:radio}.

\subsection{Centimeter: ATCA}
We observed GRB~211106A with the Australia Telescope Compact Array (ATCA) at 6 epochs via DDT project CX493 (PIs: Laskar, Bhandari, Fong), with the first epoch taken at a mid-time of 14.2~days after the burst. We used the dual-frequency, dual polarization mode of the CABB correlator, with the two IFs tuned to different frequencies to maximize spectral coverage. 
We used the 4~cm (IFs tuned to 5.5~GHz and 9.0~GHz) receiver in each epoch, and additionally observed at 15~mm (17~GHz/19~GHz) in 4 epochs and at 7~mm (33~GHz/35~GHz) in 3 epochs. We utilized PKS B1934$-$638 as bandpass and flux density calibrator and J2315$-$5018 as complex gain calibrator, except at 7~mm, where we utilized PKS B1921$-$238 as bandpass calibrator. 
The observations spanned multiple configurations. 

We analyzed the data using standard reduction procedures in Miriad, treating each IF and each epoch separately, followed by imaging in CASA with two Taylor terms, employing Briggs weighting with a robust parameter of 0. 
To improve phase coherence in the data, we generated a deep image of the field by stacking the $uv$ data from all epochs in each band separately and used the associated clean components as a model for self-calibrating the joint data set at each frequency (the target itself was not included in the model). After the second round of phase-only self-calibration, we subtracted the $uv$ model from the visibilities to generate calibrated target-only datasets. 

We combine and image the two IFs at 15~mm and at 7~mm together for maximum signal-to-noise, and report the results at the mean frequencies of 18~GHz and 34~GHz in these bands, respectively. We image 5.5~GHz and 9.0~GHz separately due to the large fractional bandwidth covered by the 4~cm receiver. We detect a radio counterpart at 5.5~GHz, 9.0~GHz, and 18~GHz at a position consistent with the mm-band position (Fig.~\ref{fig:radioimages}). There is insufficient flux in the 34~GHz images for self-calibration, and we report upper limits in this band from imaging of the field per epoch. We verify our results by performing point-source fitting directly in the $uv$ domain for each epoch and frequency band separately using \texttt{uvmodelfit} in CASA, and recover fluxes consistent within $1\sigma$ of those obtained from imaging.  
We present our ATCA flux density measurements in Table~\ref{tab:data:radio}.

\begin{figure*}
 \includegraphics[width=\textwidth]{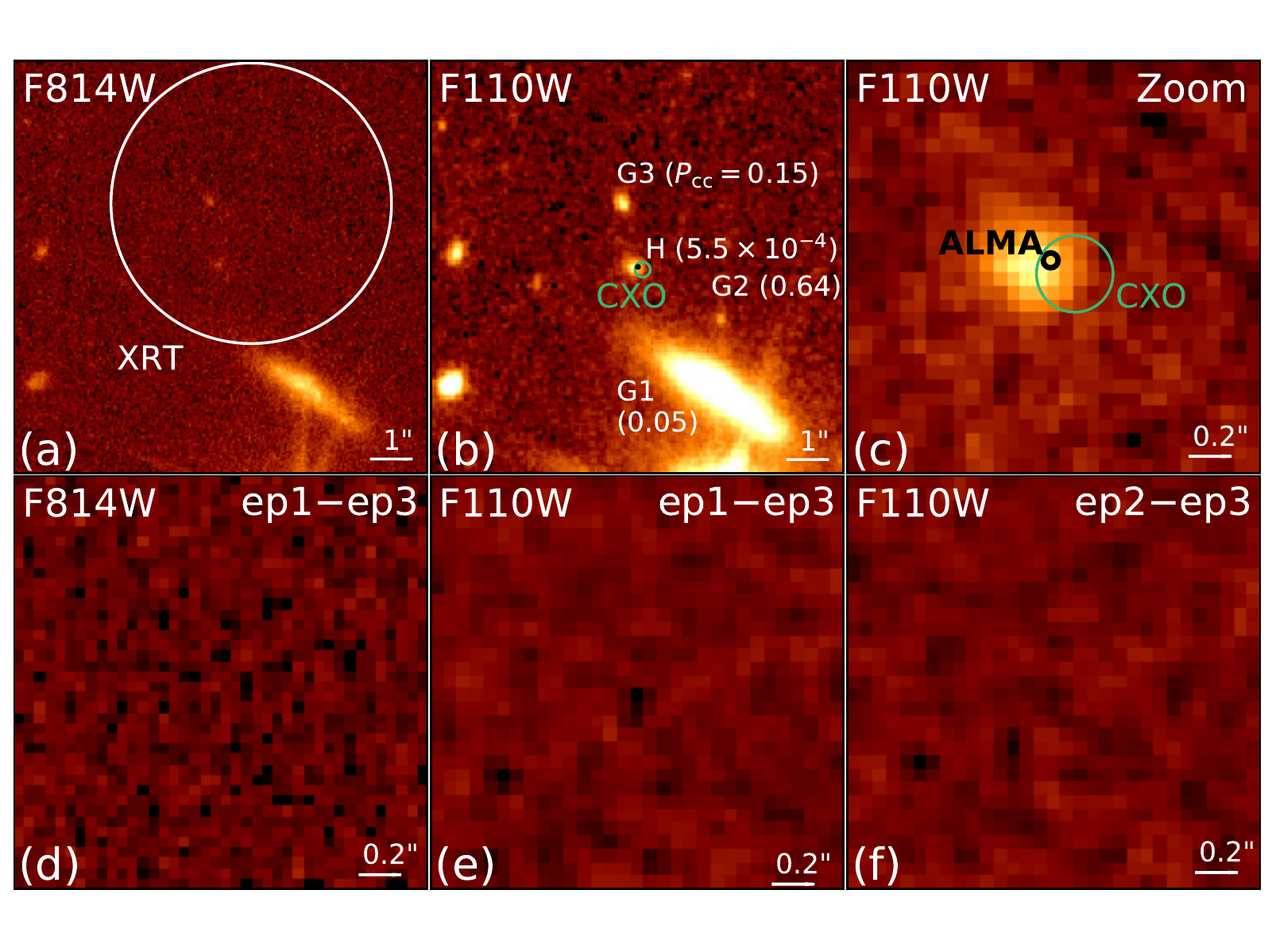}
 \vspace{-0.45in}
 \caption{\HST\ images of \grb\ from stacking all available epochs (top row) and difference imaging between epochs (bottom row). The ALMA position is coincident with a $\approx25.7$~mag optical counterpart, which has no detectable variability. (a) \HST/F814W stack with the \Swift/XRT 90\% error circle, consistent with several potential matches. (b) \HST/F110W stack with the \Chandra\ (green, registered to \HST; see Appendix~\ref{text:astrometry}) and ALMA (black) $1\sigma$ error circles. Numbers indicate probabilities of chance coincidence of objects in the image with the ALMA afterglow. (c) Zoom into the F110W stack, showing the relative offset between the most likely host galaxy and the afterglow. (d) Difference image between the two observations (ep1 \& ep3) taken in the F814W filter. (e) Difference image between the first (ep1) and third (ep3) observations in the F110W filter. (f) Difference image between the second (ep2) and third (ep3) observations in the F110W filter. No residual flux is detected in any difference image. Panels (a) and (b) are 10\arcsec on a side and panels c--f are 2\arcsec on a side. All panels are shown with the same color scale and are centered on the ALMA position, with North up and East to the left. }
\label{fig:HSTpanels}
\end{figure*}

\subsection{Optical and Near-IR: \HST}
\label{text:hst}
Optical follow-up of GRB~211106A with the VLT yielded a deep non-detection at $\approx 2.9$~days of $R \gtrsim 26.4$~mag \citep{2021GCN.31070....1M}. We observed \grb\ with the {\it Hubble Space Telescope} (\HST) over three epochs (at $\approx19.1, 25.3$, and $48.2$~days after the burst, respectively) 
with the Wide Field Camera 3 (WFC3) in F110W (all epochs) and Advanced Camera for Surveys (ACS) in the F814W band (first and third epochs only) through program 16303 (PI: Berger). We aligned and drizzled each epoch using the {\tt python}-based code {\tt hst123} \citep{hst123} as described in \citet{Kilpatrick21_gw170817}.  In addition, we drizzled all F814W and F110W images into deeper combined images and determined their absolute world coordinate system in the Gaia eDR3 frame (with rms astrometric uncertainty of 10~mas and 15~mas, respectively) using seven common astrometric standards in the \HST\ imaging and Gaia catalog \citep{gaiaedr3}.

We perform image subtraction of the first two epochs relative to the final epoch to place a limit on the afterglow emission using {\tt hotpants}
\citep{Becker15} with parameters identical to those used in \citet{Kilpatrick21_gw170817}. We perform forced photometry at the location of the mm afterglow using an empirical point-spread function (PSF) constructed in the template image frames with {\tt photutils} \citep{bsr+20}, and list the resulting upper limits in  Table~\ref{tab:data:nir}. 

\section{Host Association and Host Properties} 
\label{text:host}
To identify the most likely host galaxy, we compute the probability of chance coincidence, $P_{\rm cc}$ \citep{bkd02} of the mm-band afterglow to nearby extended objects in the \HST/F110W image and note this value next to the corresponding object in Fig.~\ref{fig:HSTpanels}. The ALMA and \Chandra\ positions are at small angular offsets of 97~mas and 211~mas, respectively, from the center of an extended object, H. We measure $m_{\rm F814W}=25.791\pm0.069$~mag and $m_{\rm F110W}=25.709\pm0.016$~mag in a 0.3\arcsec\ aperture for this object. This yields low values of $P_{\rm cc}\approx 5.5\times10^{-4}$ and $P_{\rm cc}\approx 2.6\times10^{-3}$, for the ALMA and \Chandra\ positions, respectively, where we have incorporated the localization uncertainty by combining it with the angular separation in quadrature. 

In contrast, we find much higher values of $P_{\rm cc}$ for other nearby objects\footnote{To compute observed magnitudes for the other objects, we fit the surrounding field using \texttt{galfit} and empirical PSFs, and scale the magnitudes to that of object H.}, marked G1, G2, and G3, in the \HST/F110W stack. The next-lowest value of $P_{\rm cc}\approx0.05$ is for galaxy G1 at $z=0.097$ \citep{gcn31075} at a projected separation of $9$~kpc (Fig.~\ref{fig:HSTpanels}). 
If located at the redshift of G1, object H would have an absolute magnitude of $M_H\approx-12.5$, corresponding to a luminosity, $L\approx6\times10^6\,L_\odot$, which is much greater than that of the most luminous globular clusters known \citep{rej12}. 
This rules out a globular cluster origin, and instead implies that H is a background galaxy unrelated to G1. 
If the red color ($R-F814W\gtrsim0.6$~mag) of H is due to the presence of the 4000\AA\ break between $R$-band and $F814W$, this would imply a redshift, $z\approx0.7$--1.4. 

An alternative explanation for the observed red color is a dusty stellar population. However, this is not commensurate with the relatively blue $F110W-F814W\approx-0.1$~mag color of H. To see this, we fit the observed fluxes of H with an instrinsic power law model with Small Magellanic Cloud extinction \citep{pei92}. We find both a large amount of intrinsic extinction, $A_{\rm V,H}\approx2.9$~mag and an extremely steep intrinsic spectrum, $\beta\approx1.3$ (corresponding to $F_{\lambda}\propto \lambda^{-3.3}$). This is steeper than the steepest observed UV spectral slope of local galaxies, $\beta\lesssim0.5$ \citep{wdch+11}, rendering a dusty origin of the red $R-F814W$ color highly unlikely. To further test this, we fit the observed SED of H using CIGALE \citep{nbg+09} at four different redshifts ($z=0.097$, $z=0.5$, $z=1$, and $z=2$; see Appendix \ref{text:cigale}). While the fit at $z=1$ is able to account for the red $R-F814W$ color by ascribing the flux decrement to the 4000\AA\ break, fits at the other redshifts are systematically worse. This supports the hypothesis that H is unrelated to G1 and is at a moderately high redshift, $z\approx1$.

The observed F110W magnitude of object H corresponds to an absolute magnitude of
$M\approx-16.2$ ($L\approx2\times10^8L_\odot\approx0.02L_*$, roughly rest-frame $I$-band) at $z=0.5$ and
$M\approx-17.7$ ($L\approx10^9L_\odot\approx0.05L_*$, roughly rest-frame $V$-band) at $z=1$ (without $K$-corrections for SED shape). Even without accounting for color corrections, this would place H at the low-luminosity end of the SGRB host luminosity function \citep{ber14}. Alleviating this by supposing a redshift of $z\gtrsim1$ would imply even more extreme properties for the afterglow (Section~\ref{text:luminosity}).  
Using the empirical PSFs derived from the image to fit H with an elliptical Gaussian model using \texttt{galfit} \citep{phir02}, we obtain a full-width at half-maximum (FWHM) of $0.260\pm0.008$~arcsec and axis ratio, $b/a=0.73\pm0.03$ at position angle, $\theta_{\rm PA}=68\pm4$~deg. Normalized to the host effective radius of $\sigma_r={\rm FWHM}/2.354=110\pm3$~mas, the offset of the ALMA afterglow is $\approx0.9\sigma_r$. This corresponds to a physical separation of $\approx0.6$~kpc and $\approx0.8$~kpc at $z=0.5$ and $z=1$, respectively, placing \grb\ at the extreme lower end of the median SGRB offset distribution, both in terms of physical and host-normalized offsets \citep{ber14}. 

\begin{deluxetable*}{ccccccccccc}
    \label{tab:params}
 \tabletypesize{\footnotesize}
 \tablecolumns{9}
 \tablecaption{Afterglow model parameters from multi-wavelength modelling of \grb}
 \tablehead{   
   \colhead{$z$} &
   \colhead{IC/KN} &
   \colhead{$p$} &
   \colhead{$\log\epse$} &
   \colhead{$\log\epsb$} &
   \colhead{$\log\dens$} &
   \colhead{$\log\EKiso$} &
   \colhead{\tjet} &
   \colhead{$\thetajet$} &
   \colhead{$A_V$} &
   \colhead{$\EK$} 
   }
\startdata
     1.0 & Y & $2.47\pm0.05$ &  $-0.08^{+0.06}_{-0.11}$ & $-5.04^{+0.80}_{-0.66}$ & $-0.59^{+0.18}_{-0.20}$ & $53.22^{+0.31}_{-0.34}$ & $29.23^{+4.53}_{-4.01}$ & $15.51\pm1.43$ & $4.95^{+2.05}_{-1.47}$ & $51.79^{+0.27}_{-0.30}$\\
     0.5 & Y & $2.59\pm0.04$ & $-0.06^{+0.04}_{-0.13}$ & $-4.86^{+0.62}_{-0.58}$ & $-0.93^{+0.20}_{-0.46}$ & $52.72^{+0.47}_{-0.27}$ & $31.71^{+5.69}_{-4.57}$ & $18.56^{+1.61}_{-3.28}$ & $5.62^{+1.66}_{-1.42}$ & $51.42^{+0.27}_{-0.24}$\\
     1.0 & N & $2.19^{+0.06}_{-0.05}$ & $-0.71^{+0.13}_{-0.16}$ & $-0.21^{+0.09}_{-0.16}$ & $-2.19^{+0.34}_{-0.46}$ & $51.69^{+0.21}_{-0.15}$ & $32.10^{+4.68}_{-3.99}$ & $15.70^{+2.12}_{-2.57}$ & $5.28^{+1.89}_{-1.35}$ & $50.26^{+0.08}_{-0.07}$\\
     0.5 & N & $2.63\pm0.03$ & $-0.84^{+0.51}_{-0.67}$ & $-3.63^{+2.50}_{-2.17}$ & $-4.92^{+2.00}_{-2.91}$ & $54.30^{+0.30}_{-0.54}$ & $39.28^{+4.14}_{-3.55}$ & $4.18^{+3.46}_{-2.46}$ & $5.15^{+1.92}_{-1.47}$ & $51.54\pm0.58$\\
\enddata
\tablecomments{Units are as follows: $\dens$ is in $cm^{-3}$, $\EKiso$ and $\EK$ are in erg, \tjet\ is in days, $\thetajet$ is in degrees, and $A_V$ is in mag.}
\end{deluxetable*}

\section{Multiwavelength Modeling}
\label{text:modeling}
We now turn to an analysis of the extensive afterglow data. We interpret the observed X-ray to radio observations in the context of synchrotron radiation from an afterglow forward shock (FS) produced by the interaction of the relativistic GRB jet with its environment \citep{spn98,gs02}. We assume a uniform density (ISM) environment (as expected for a compact binary progenitor) and a particle acceleration fraction, $f_{\rm NT}=1$ \citep{ew05,rl17}. The parameters of this model are the isotropic-equivalent energy release (\EKiso), density of the environment (\dens), the fraction of the shock energy given to relativistic electrons (\epse) with energy power-law index, $p$, and the fraction imparted to magnetic fields (\epsb). The resulting spectrum is characterized by three break frequencies: the self-absorption break (\nua), the characteristic synchrotron frequency (\numax), and the cooling break (\nuc). We include inverse-Compton (IC) cooling effects on the synchrotron spectrum, together with Klein-Nishina (KN) corrections \citep{se01,nas09,jbvdh21}.

\begin{figure*}
  \centering
  \includegraphics[width=0.95\textwidth]{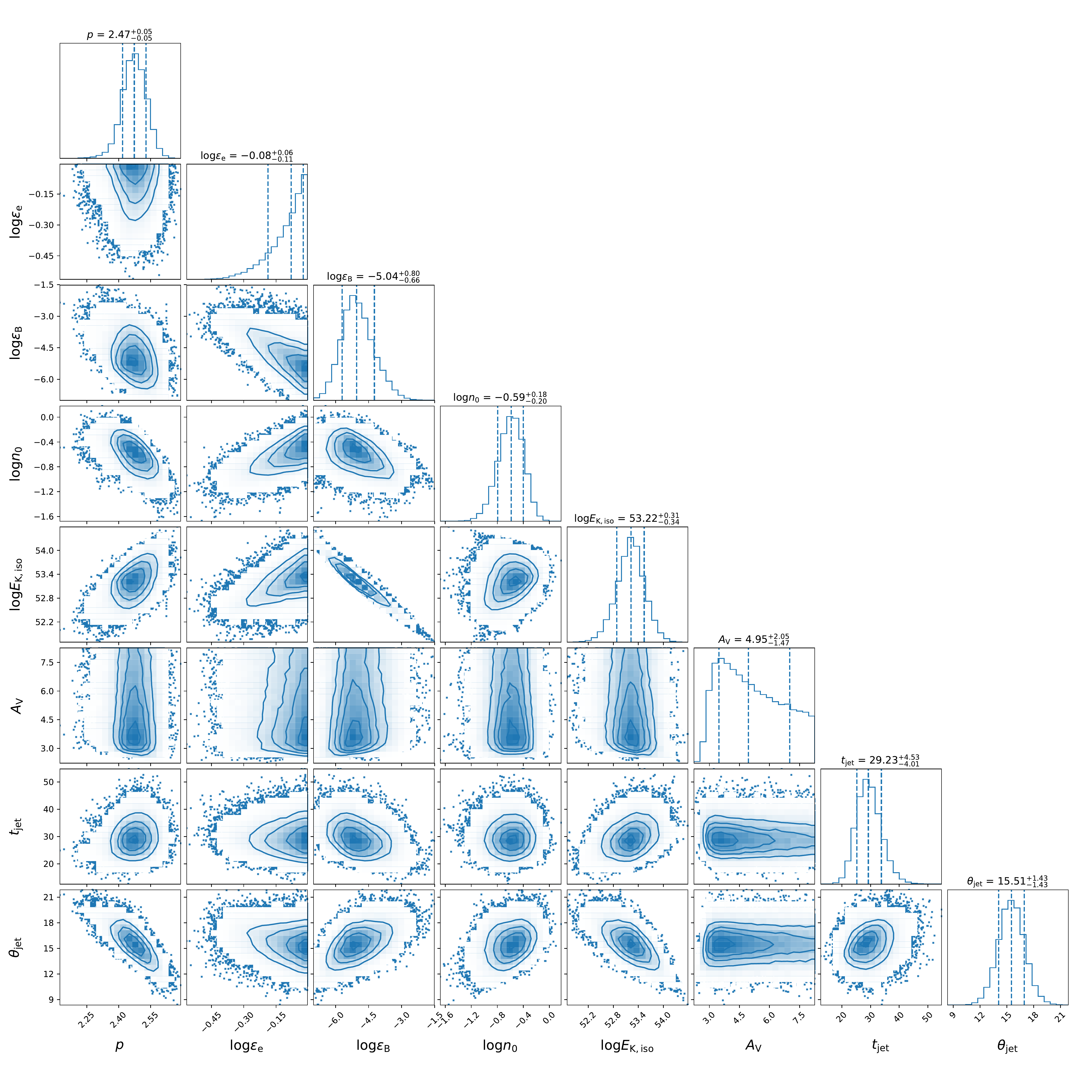}
  \vspace{-0.2in}
  \caption{Correlations and marginalized posterior density from multi-wavelength modeling of \grb\ at $z=1$ (including IC/KN effects), with \dens\ in \pcc, \EKiso\ in erg, \tjet\ in days, and the opening angle ($\thetajet$) in degrees. \thetajet\ is derived from \EKiso, \dens, and \tjet\ \citep{sph99} and is not an independent free parameter. The contours enclose 39.3\%, 86.4\% and 98.9\% of the probability mass in each correlation plot (corresponding to $1\sigma$, $2\sigma$, and $3\sigma$ regions for two-dimensional Gaussian distributions), while the dashed lines in the histograms indicate the 15.9\%, 50\% (median), and 84.1\% quantiles (corresponding to $\pm1\sigma$ for one-dimensional Gaussian distributions). See Table~\ref{tab:params} for a summary.}
\label{fig:corner}
\end{figure*}

\subsection{Preliminary Considerations}
\label{text:prelim}
The X-ray light curve can be fit as a single power law\footnote{We employ the convention $F_{\nu}\propto t{^\alpha}\nu^{\beta}$ throughout.}, with decline rate $\alpha_{\rm X}=-0.97\pm0.03$ (Fig.~\ref{fig:Xraylc}), which would imply $p\approx1.9$ if $\nuc<\nux$ and $p\approx2.3$ if $\nux<\nuc$ under the standard synchrotron framework (ignoring IC/KN effects). The expected spectral index in these cases is $\beta\approx-0.9$ and $\beta\approx-0.6$, respectively, both of which are consistent with the observed X-ray spectral index of $\beta_{\rm X}=-0.92\pm0.30$. The X-ray data then suggests $p\approx1.9$--2.3, but does not yield a definitive constraint on the location of $\nuc$. 

The ALMA 97.5~GHz light curve appears flat ($F_{\nu,3mm}\approx0.15$~mJy) from $\approx13$ to $\approx20$~days, after which it declines steeply as $\alpha_{\rm 3mm}\approx-1.5$ (Fig.~\ref{fig:mm_cm_comparison}).
The shallow light curve before the break indicates that the spectral peak ($\numax$) passes through the 3~mm band at $\approx13-20$~days with a flux density, $\fnumax\approx0.15$~mJy. The steepest decay at $\numax\lesssim\nu\lesssim\nuc$ is expected to be $\alpha\approx3(1-p)/4\approx-0.9$ for $p\approx2.2$. Thus, unless there is a change in $p$, or it is much higher (i.e., $p\approx3$) than estimated from the X-ray light curve ($p\approx2$--2.3), the rapid post-break decline suggests a jet break prior to the last ALMA detection at $\approx43$~days.

A broken power law fit to the ATCA C-band (5.5~GHz) data yields a rise rate $\alpha_{C,1}=0.26\pm0.10$, a decline rate, $\alpha_{C,2}=-2.4\pm0.8$, break time $56\pm4$~days, and peak flux density, $F_{\nu,C,max}=0.14\pm0.01$~mJy. The fact that the 5.5~GHz light curve does not decline appreciably until $\gtrsim42$~d, whereas the ALMA light curve starts declining much earlier at $\lesssim28$~days is consistent with the interpretation of a jet break at $\lesssim43$~days, and with light curve turn-over in the radio/mm bands arising from the cascading passage of $\numax$ through the mm/cm bands. 

Interpolating the X-ray light curve to the time of the VLT upper limit at $\approx2.9$~days, we find an X-ray to optical spectral index of $\beta_{\rm OX}\gtrsim -0.39$. This indicates that the optical flux is strongly suppressed relative to the expectation from the standard synchrotron model ($\beta_{\rm OX}>-0.5$). Furthermore, the observed X-ray spectral index $\beta_{\rm X} \approx -0.92$ implies $\beta_{\rm OX} - \beta_{\rm X} \gtrsim 0.52$, and thus \grb\ satisfies the definition of a dark burst of both \citet{jhf2004} and \citet{vkg2009}. Several other SGRBs have been classified as dark \citep{bcfc09,fbm+12,bzl2013}, and we account for the dark nature by incorporating host extinction in our analysis using an SMC extinction model \citep{pei92}.

Finally, our {\it HST} limits at $\gtrsim19$~days cannot be used to place meaningful constraints on an AT2017gfo-like kilonova or previous SGRB kilonova candidates. The VLT limit at $2.9$~days only probes to depths of comparable to $\approx10$ times the luminosity of AT\,2017gfo for an assumed redshift of $z=0.5$, while no meaningful constraints on kilonova emission can be derived from these optical/NIR observations at $z=1.0$. 

\subsection{MCMC Modeling}
\label{text:mcmc}
We now search the parameter space of $p$, \EKiso, \dens, \epse, \epsb, \tjet, and \AV\ for the best-fit synchrotron model to the afterglow observations using Markov Chain Monte Carlo (MCMC) with \texttt{emcee} \citep{fhlg13}. The details of our implementation are described in \citet{lbz+13,lbt+14}. We include the effects of KN corrections for the first time (McCarthy \& Laskar in prep) using the prescription of \cite{nas09} as described by \cite{jbvdh21}. We run 512 walkers for 2000 steps, discarding an initial period of 30-200 steps (judged by stationarity in the resulting posterior density function) as burn-in. We use a uniform prior on $p$ from 2.001 to 2.99 and on the intrinsic extinction, $A_V\lesssim8$~mag. We restrict $\log(\epse)$ and $\log(\epsb)$ to the range $\in(-7,0)$ with the additional constraint $\epse+\epsb<1$. We constrain $\log(\dens)\in(-10,10)$ and $\EKiso\in(10^{48},5\times10^{54})$. We use \cite{jef46} priors for these last four parameters. We also perform the analysis without IC/KN corrections in each case for comparison, resulting in a total of four sets of parameters. We summarize the results of our MCMC analysis in Table~\ref{tab:params}.

\begin{figure*}
 \centering
 \includegraphics[width=\columnwidth]{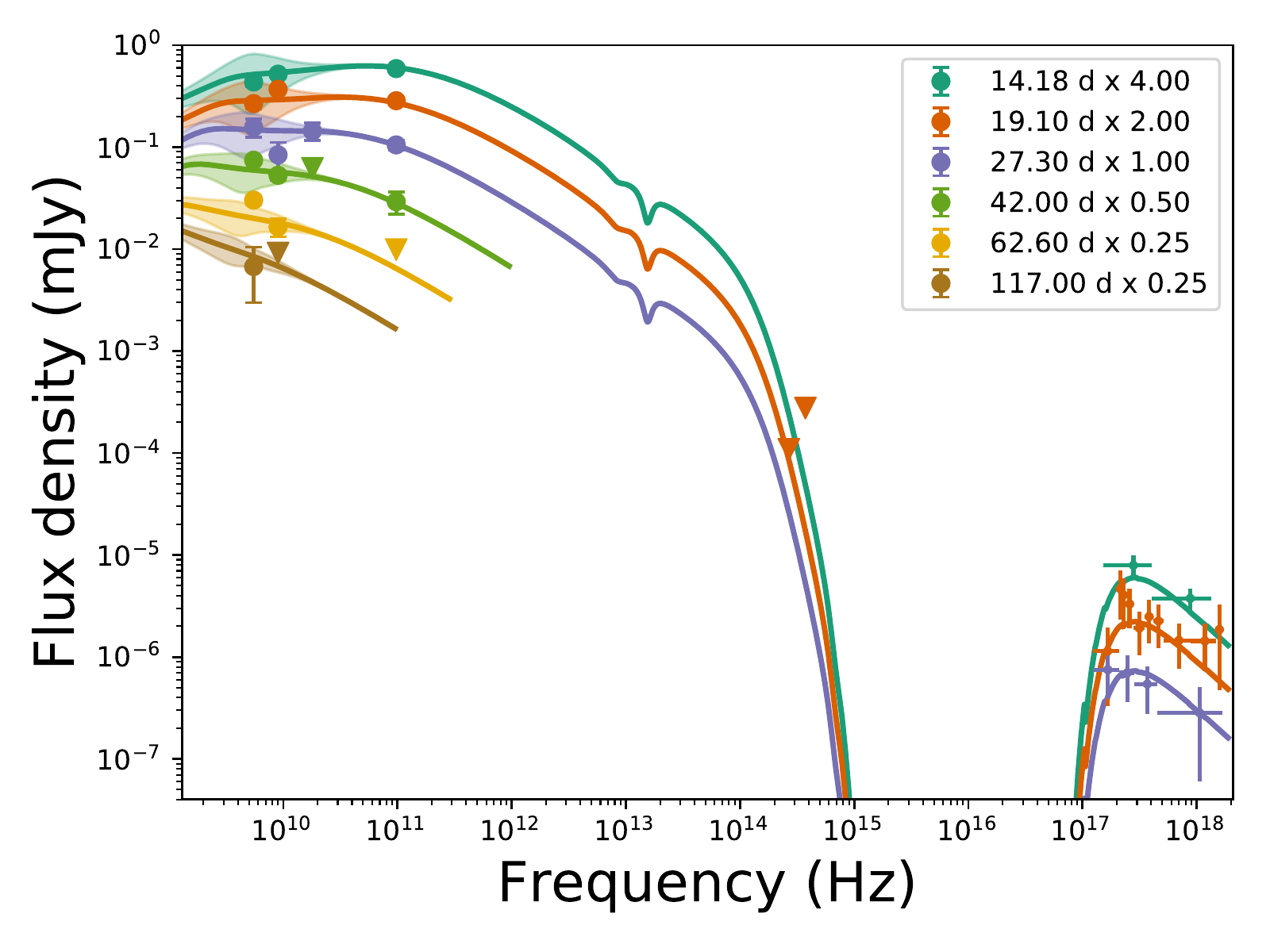}
 \includegraphics[width=\columnwidth]{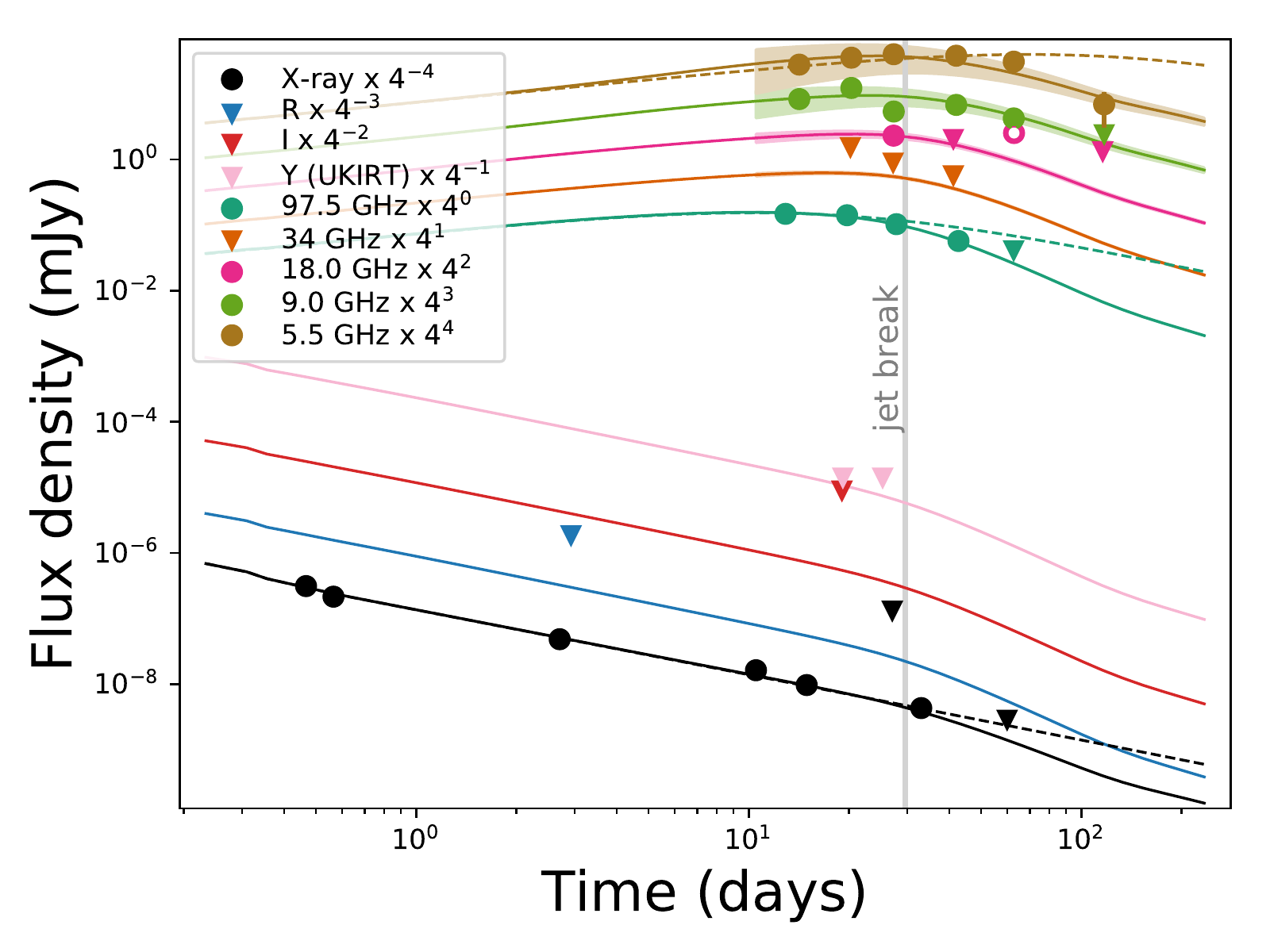}
 \caption{{\it Left}: Spectral energy distributions of \grb\ afterglow from radio (circles) to X-rays (crosses) spanning from 14.18 to 117~days, together with a best-fit (highest-likelihood) forward shock ISM model (lines) at $z=1.0$, including expected contribution from interstellar scintillation (shaded bands). Triangles indicate upper limits. We have interpolated the observations, where necessary, to the common times for each epoch using broken power law fits to the ALMA/ATCA 5.5~GHz light curves. Upper limits are not interpolated. The X-ray spectra have been scaled to the times of the SEDs using a broken power law fit to the X-ray light curve. The \HST\ upper limits require $\AV\gtrsim2.6$~mag of extinction in the host galaxy. {\it Right}: Corresponding light curves with (solid) and without (dashed) a jet break. The model reproduces all observations, except for the 18~GHz detection at 62.7~days, which is masked during modeling (open circle). The turnover in the mm-band light curve and the steep decline in the cm-band light curve at $\gtrsim62.7$~days require a jet break at $\tjet\approx29$~days (grey, vertical line), constraining the jet opening angle to $\thetajet\approx16$~degrees. }
\label{fig:ISMfit}
\end{figure*}

We find that some of the derived parameters are sensitive to the choice of redshift. The cooling frequency is between the optical and X-rays in the $z=1$ models, above the X-rays in the $z=0.5$ model without IC/KN and $\nuc\approx\nux$ in the $z=0.5$ model when including IC/KN. In addition, the parameters derived from including IC/KN effects are quite different from those without. We note that the fits achieved by turning off IC effects yield Compton-$Y$ parameters at $\nuc$ of $Y_{\rm c}<1$, where IC effects are indeed negligible (and vice-versa), and thus all four sets of parameters are internally self-consistent. However, the fits without IC corrections are slightly systematically worse (maximum log likelihood of $\mathcal{L}=99.97$ and 100.39 for the $z=0.5$ and $z=1.0$ fits, respectively) compared to the fits including IC/KN effects ($\mathcal{L}=103.94$ and 108.22 for $z=0.5$ and $z=1.0$, respectively). Since the number of parameters in the models are the same, models with higher likelihoods are slightly statistically favored. Given the moderately high redshift of $z\approx1$ favored by the host SED, we focus the rest of the discussion to the $z=1.0$ IC/KN model, with the understanding that some of the numerical results, in particular, are sensitive to these choices. We discuss the impact of the KN corrections in Appendix~\ref{text:KN}. For completeness, we present the $z=1.0$ model without IC/KN effects in Appendix~\ref{text:noKNmodel} and include it in parameter comparisons below, where relevant.

For our fiducial parameter set ($z=1$ with IC/KN corrections), we plot the correlation contours and marginalized posterior density functions in Fig.~\ref{fig:corner}. 
The parameters of the highest-likelihood model are, $\EKiso\approx1.9\times10^{53}$~erg, $\dens\approx0.5~\pcc$, $\epse\approx0.97$, $\epsb\approx5\times10^{-6}$, and $p\approx2.4$.
For this model, $\numax$ passes through the ALMA 3mm band at $\approx18$~days with a flux density at $\numax$ of $\fnumax\approx0.15$~mJy, which is consistent with the constraints from the ALMA light curve. We also estimate a jet break time of $\tjet\approx29$~days, consistent with the steepening observed in the cm- and mm-band light curves. We note that whereas we have provided ranges for the parameter $A_{\rm V}$, this parameter is unbounded above, since there was no detection of an optical transient associated with the event. However, we can establish a lower limit of $A_{\rm V}\gtrsim2.6$~mag, corresponding to the value above which 99.7\% of the MCMC samples reside. Our derived values of $\AV$ are consistent with the $\AV$--$N_{\rm H,int}$ correlations for dark GRBs \citep{zbm+13}.  We plot our model light curves and SEDs for the highest likelihood parameter set in Fig.~\ref{fig:ISMfit}. 

Finally, we derive a very high value of $\epse$ for both models upon inclusion of IC/KN effects. We note that the allowed range of $\epse$ spans a factor of $\approx2$ and furthermore this (and indeed all derived parameters) are degenerate with respect to the unknown electron participation fraction, $f_{\rm NT}$ \citep{ew05}. A value of $f_{\rm NT}\approx0.1$, as estimated from particle-in-cell simulations, would alleviate this issue by a corresponding factor \citep{ss11}. Capturing emission or absorption from thermal electrons would resolve this degeneracy \citep{rl17}. 

\section{Discussion}
\label{text:discussion}

\subsection{The jet opening angle}
\label{text:thetajet}
The ALMA mm-band observations of \grb\ were vital for constraining the jet break time and to derive the beaming corrected energy unencumbered by scintillation effects in the cm-band and complications from IC/KN corrections in the X-ray band. This contribution is especially important in this case due to the absence of detectable optical afterglow emission, and since the jet break occurs after the X-ray afterglow has faded beyond the sensitivity of \Chandra. The identification of the jet break, in combination with measurements of the circumburst density and energy for this burst, allows us to constrain the jet opening angle to $\thetajet\approx16^\circ$, and this value appears relatively robust to the modeling uncertainties discussed above. The one notable exception in the case of the $z=0.5$ model without IC corrections is driven by the extremely low density and high $\EKiso$, which itself arises from a strong degeneracy between these parameters for this model\footnote{In this model, $\nuc>\nux$ and is unconstrained, resulting in additional model parameter degeneracies.}. We find that removing the mm-band data from the fit and re-running the MCMC results in similar degeneracies, further highlighting the importance of securing mm-band detections. 

Eight other SGRBs have robust opening angle measurements\footnote{We exclude GRB~150424A, for which the reported opening angle assumes values for both \dens\ and \EKiso\ \citep{jlw+18}.} from identification of jet breaks in their light curves, with measured values of \thetajet\ spanning from 1--14 degrees \citep{sdpc07,sbk+06,nkr+11,fbm+12,fbm+14,tsc+16,ltl+19,tcbg+19,flr+21,ctdc21}.
Whereas an additional 5 events have robust lower limits on \thetajet\ (i.e., incorporating their \EKiso\ and \dens), only one of these has a larger inferred lower limit than this ($\thetajet\gtrsim25^{\circ}$ for GRB~050724A; \citealt{Grupe2006}). 
Thus, the opening angle for \grb\ is one of the widest inferred for SGRBs, and the resulting late jet break ($\tjet\approx29$~days) is the latest observed in any SGRB. This confirms the finding of \citet{fbmz15} that afterglow observations at $\gtrsim25$~days are essential for obtaining strong constraints on \thetajet. This late jet break ensures that the mm-band afterglow remains detectable for longer. We discuss detectability of mm-band afterglows further in Section~\ref{text:detectability}. 

\subsection{Afterglow luminosity and energetics}
\label{text:luminosity}
We find that the afterglow of \grb\ has several superlative properties. 
In comparison with the population of \Swift/XRT SGRB afterglows, the X-ray afterglow of \grb\ is one of the most luminous at a comparable rest-frame time (Fig.~\ref{fig:Xraylc}). 
Similarly, the cm afterglow of \grb\ is extremely long-lived, and, at $z=1.0$, is more luminous than any other SGRB radio afterglow (second-most luminous if at $z=0.5$). The luminosity of this mm-band afterglow rivals that of several LGRBs (Fig.~\ref{fig:mm_cm_comparison}). 
These properties are reflected in the high $\EKiso\approx1.6\times10^{53}$~erg (median value from the MCMC) in our fiducial model ($z=1$ with IC/KN corrections), which is two orders of magnitude larger than the median values of $\EKiso\approx(1$--$3)\times10^{51}$~erg inferred for the SGRB population \citep{fbmz15}.
This yields a prompt $\gamma$-ray efficiency of $\eta_\gamma\equiv\Egammaiso/(\EKiso+\Egammaiso)\approx0.03$ for the $z=1.0$ model ($\approx0.02$ at $z=0.5$). This is the second-lowest prompt efficiency inferred for SGRBs after GRB~150101B with $\eta_\gamma\approx10^{-3}$, but consistent with the wide range spanned by this parameter for SGRBs \citep{fbmz15}. 

The true (beaming-corrected) energy is also high, with median values from the MCMC (in units of erg) of $\log\EK\approx51.8$ and $50.3$ for the $z=1$ model with and without IC/KN corrections, respectively. To put this in context, we compute $\EK$ for all SGRBs that have published values of $\EKiso$ and either measurements or lower limits on $\thetajet$, resulting in a sample of 12 events from \cite{fbmz15} and 4 additional events published subsequently \citep{jlw+18,ltl+19,tcbg+19,rfv+21,flr+21}. We find that the highest value of \EK\ was obtained for GRB~180418A ($\log\EK\gtrsim51.7$; \citealt{rfv+21}). Two additional events have $\log\EK$ values between those of the two $z=1$ models (GRBs~120804 and 140930B, with $\log\EK\gtrsim50.8$ and $\gtrsim50.3$, respectively). The remaining 13 events all $\log\EK<50.3$. This places \grb\ in the top $\gtrsim80\%$ of SGRBs with measured beaming-corrected energy, making it one of the most energetic SGRBs to date. The mm-band detection was pivotal in this measurement, as it is the only band that samples both $\numax$ and $\fnumax$ prior to the jet break, thus breaking the \EKiso--\dens\ degeneracy.

\subsection{\grb\ and the detectability of mm-band SGRB afterglows}

\begin{figure*}
 \centering
 \includegraphics[width=\textwidth]{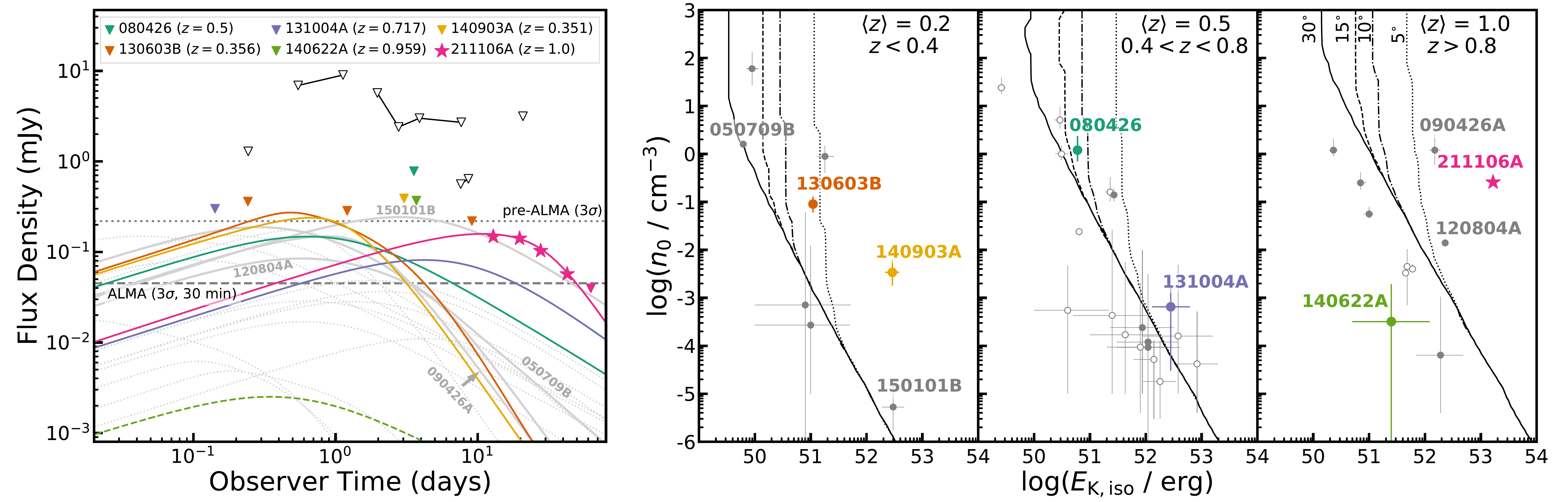}
 \caption{{\it Left}: Theoretical 90~GHz light curves for 22 SGRBs (plus 211106A, magenta stars) with known redshift and published \EKiso\ and \dens\ (lines, including collimation effects, where a jet break has been inferred). Six events have mm-band observations (colored points and lines). Empty triangles are mm-band upper limits for an additional 5 SGRBs with unknown physical parameters. 
 Previous observations have failed to detect mm-band afterglows owing to a combination of insufficient sensitivity (pre-ALMA; horizontal dotted line) and incommensurate cadence (Section~\ref{text:detectability}); however, 
 9/23 events (indicated by solid lines) would have been detected by ALMA in 30 minutes on-source (horizontal dashed line). {\it Right}: \dens\ vs \EKiso\ for SGRB afterglows from \citet{fbmz15} divided into three redshift bins, with the median redshift and redshift range indicated on the upper right corners. Events that have been observed in the mm-band are indicated by colored circles and those with no measured redshifts are plotted as open circles at fiducial values of $z=0.5$ or $z=1.0$. The position of \grb\ is marked by the magenta star. The parameter space in each sub-plot is divided by lines at four different values of \thetajet\ (values indicated on the right-most sub-panel), to the right of which the afterglow is expected to be detectable with ALMA for $\gtrsim2$~days (Section~\ref{text:detectability}). }
\label{fig:detectability}
\end{figure*}

\label{text:detectability}
It is reasonable to ask whether an unprecented value of some physical property (e.g.~ high \EKiso, large opening angle) for \grb\ places it in a position in parameter space that makes this event uniquely suitable for detection in the mm-band, or whether the improvement in sensitivity in the mm-band with the advent of ALMA would have soon yielded such a discovery for a SGRB afterglow anyway. Alternatively, perhaps we simply missed previous mm afterglows because we did not observe them at the right time, and the relatively late commencement of the mm-band follow-up in this case coincidentally yielded just the right temporal sampling of the light curve? We now address these questions, beginning by investigating the mm-band light curves of all SGRBs with published mm-band upper limits in the context of the synchrotron model. 

A total of 11 SGRBs have been observed in the mm-band so far (Fig.~\ref{fig:mm_cm_comparison}). Of these, three events have no afterglow detection at any wavelength (020531, 051105A, 140606A), and an additional two (GRB~050509B and GRB~060801) do not have data\footnote{The X-ray light curve of GRB~050509B is poorly sampled and that of GRB~060801 is dominated by an initial steep decay. Neither event was detected at any other wavelength.} of quality sufficient for constraining physical parameters. In Fig.~\ref{fig:detectability}, we plot the mm-band observations for the remaining six SGRBs: GRB~080426 \citep{duplm+12}, 130603B, 140622A, 140903A \citep{phct+19}, 131004A \citep{fbmz15}, and \grb\ (this work), along with model light curves corresponding to published values of the physical parameters for each event. Where multiple sets of physical parameters are available (e.g., for two different assumed values of \epsb), we plot the more optimistic model (except in the case of GRB~130603B, as discussed below), which is always the one with lower assumed \epsb\ and higher inferred density. 

In the case of GRBs~080426, 131004A, and 140622A, we find that the peak flux of the mm-band light curve is below the only published upper limits for these events by factors of $\approx5.3$, 3.7, and 147, respectively. Two of these events have extremely low values of density ($\dens\approx6.5\times10^{-4}$~\pcc\ for GRB~131004A, and $\dens\approx3.2\times10^{-4}$~\pcc\ for 140622A). Although the third (GRB~080426) has higher density ($\dens\approx1.2$~\pcc) it also has one of the lowest inferred energies for SGRBs ($\EKiso\approx6\times10^{50}$~erg). For $p\approx2.2$ and at the typical SGRB redshift of $z\approx0.5$, the spectral peak flux density is given by, $\fnumax\approx40(\epsb/10^{-2})^{1/2}(\dens/10^{-2}~\pcc)E_{\rm K,iso,51}~\mu$Jy \citep{gs02}. This implies that SGRBs with $\dens\lesssim10^{-2}$~\pcc\ or $\EKiso\lesssim10^{50}$~erg are unlikely to be detectable with both past and present mm-band facilities, and confirms that these three previous events evaded detection due to their lower density or energy. 

This leaves two events: GRBs~130603B and 140903A. Both the energy and density of GRB~130603B are higher than the above thresholds. For GRB~140903A, while the inferred density is low ($\dens\approx3.4\times10^{-3}$~\pcc), the energy (which has a stronger impact on \fnumax) is high ($\EKiso\approx3\times10^{52}$~erg) and thus both these events should have been detectable by the metric of peak flux density. For these two events, the reason for mm-band non-detections appears to be their narrow collimation angles, $\thetajet\approx6^\circ$ and $\thetajet\approx4^\circ$, respectively. Physically, narrower collimation corresponds to a lower true energy, \EK. Observationally, the earlier jet-break limits the peak flux of the mm-band light curve. In the case of GRB~140903A, the mm model light curve peaks at $\approx0.24$~mJy, which is lower than the spectral peak flux prior to the jet break ($\fnumax\approx0.5$~mJy), and below the PdBI upper limit of $\approx0.4$~mJy, thus explaining the mm-band non-detection. Two models are available for GRB~130603B with different values of \epsb\ \citep{phct+19}. The model with lower \epsb\ and higher density actually over-predicts the existing mm-band observations (even upon including the jet break) and we can rule this model out. While the lower density model \textit{does} produce fluxes higher than the deepest upper limits for this event, the timing and depth of the epochs unfortunately do not probe the underlying light curve. The early observations were not deep enough and by the time deeper observations were taken, the light curve would have faded below detectability due to the early jet break. This suggests that early, deep observations are essential to capture the mm counterparts of narrowly collimated / low-\EK\ outflows. 

The inferred energy and density for \grb\ (in all models) is higher than the thresholds discussed above. The wide opening angle and resulting late jet break ($\tjet\approx29$~days) further drive the long-lived mm-band afterglow. Finally, the factor of 5--10 higher sensitivity of ALMA compared to CARMA and PdBI has further broadened the detectability window. To illustrate this, we plot model light curves for all 17 SGRBs at known redshifts with published \EKiso\ and \dens\ values (but without mm observations) as grey lines in the left panel of Fig.~\ref{fig:detectability}. We find that the mm-band afterglows of 9/23 (39\%) SGRBs (solid lines) would have been detectable for at least $\gtrsim2$~days (observer frame) with ALMA, while only one event (GRB~150101B) satisfies this condition at pre-ALMA sensitivity levels.

To further quantify this, we compute the duration for which SGRB mm-band afterglows are detectable with ALMA at $3\sigma$ in 30~min of on-source integration time ($F_{\nu}\gtrsim50~\mu$Jy) at 90~GHz for different values of \dens, \EKiso, and \thetajet, and compare the results with the inferred parameters for a sample of 38 events from Table 3 of \cite{fbmz15}, which forms an X-ray complete parent sample spanning 10 years. We plot the results, divided into three redshift bins, in the right panel of Fig.~\ref{fig:detectability}. The mm-band afterglows of the events to the right of the lines (drawn for four different jet opening angles) are detectable with ALMA for more than 2 days. We find that, independent of the opening angle, \grb\ would have been detectable owing to its position in the \dens-\EKiso\ space alone. Its wide jet (and hence, high \EK) further ensured a high likelihood of discovery upon triggering of mm-band observations. 

The detectability of the other events is contingent on their unknown opening angle (or, equivalently, their unknown true \EK), although some events (especially at high redshift) simply cannot be detected owing to a combination of low density and/or energy, as previously suggested. 9 events fall to the right of the $\thetajet=5^\circ$ line, and these events, even if narrowly collimated (i.e., with low \EK), would have been detectable with ALMA. On the other hand, 22 events (58\%) would not have been detectable for any value of their intrinsic \thetajet\ or \EK. 

If we assume that the sample of 38 events in \cite{fbmz15} is representative of the SGRB population, then if SGRBs with X-ray afterglows were to be uniformly followed up in the mm band, we might expect a conservative success rate of $r_{\rm det}\approx9/38\approx24\%$ (corresponding to the 9 events that fall to the right of the $\thetajet\lesssim5^{\circ}$ lines) and a detection rate of $r_{\rm mm}\approx 9/10\approx0.9$ mm afterglows per year (as the sample spans 10 years). For events with wider jets $\thetajet\gtrsim30^\circ$, the corresponding rates are $r_{\rm det}\approx16/38\approx42\%$ and $r_{\rm mm}\approx1.6$ per year. These rates are even better than the discovery rates ($\approx7\%$) of SGRB afterglows in the cm-band \citep{fbmz15}. At pre-ALMA levels, the mm-band detection rate is poorer by a factor of $\approx3$ with $r_{\rm det}\approx6/38\approx16\%$. 
We conclude that all three aspects (high density, wide opening angle / higher \EK, and improved sensitivity) have contributed to the discovery of the mm-band afterglow of \grb. Systematic ALMA follow-up of SGRBs should yield a significant (24--40\%) discovery rate of mm-band afterglows, potentially outpacing cm-band detections. 

\section{Conclusions}
We have presented ATCA, ALMA, \HST, \XMM, \Chandra, \Swift/XRT, \Swift/BAT, and \Fermi-GBM observations of \grb. Our $\gamma$-ray temporal and spectral analysis confirms this event as a bona-fide short-duration GRB with exceptional afterglow properties. Our ALMA mm-band detection localizes the afterglow to a faint host galaxy at $0.7\lesssim z\lesssim1.4$.
A comparison of the X-ray, mm, and radio light curves of the afterglow to that of the SGRB population reveals that this event likely possessed one of the most luminous SGRB afterglows at all these bands to date. 
The lack of an optical counterpart to deep limits implies a dust-obscured burst with an extinction, $A_{\rm V}\gtrsim2.6$~mag, further consistent with the high intrinsic X-ray absorption column density. 

We have presented the first mm-band afterglow detection of a short-duration GRB. Our well-sampled ALMA 97.5~GHz light curve for this event allows us to constrain the spectral peak frequency, peak flux density, and jet break time. We find a jet opening angle of $\thetajet\approx16^{\circ}$, the largest yet measured for an SGRB, and the resultant beaming-corrected kinetic energy, $\EK\approx2\times10^{50}$--$6\times10^{51}$~erg, is among the largest yet inferred for SGRBs. We conclude that the combination of high energy and high density, together with the improvement in sensitivity offered by ALMA, all contributed to the detection of this afterglow in the mm band. We find that a larger fraction ($\approx40\%$) of GRBs with known redshifts will be detectable with ALMA (compared to $\lesssim16\%$ with pre-ALMA facilities), but that the population will likely still be dominated by energetic events ($\EKiso\gtrsim10^{50}$~erg) in high-density ($\dens\gtrsim10^{-2}$~\pcc) environments. However, exceptions are possible for nearby ($z\lesssim0.5$) events. 

The rapid triggering and archival of BAT data by the GUANO system enabled a prompt localization and afterglow follow-up for this event, underscoring the importance of rapid-response, software-based implementations for enhancing target-of-opportunity science with time-domain observatories such as \Swift. The discovery of the cm/mm-band counterpart $\gtrsim12$~days after the trigger highlights the importance of sustained, deep radio follow-up of short-duration GRBs. 
The unusual energetics and host properties of \grb\ suggest that there may be an even greater diversity in SGRB properties than currently known, necessitating continued identification, classification, and multi-wavelength follow-up of these extreme events. 

\section{Acknowledgements}
TL acknowledges support from the Radboud Excellence Initiative. The Fong Group at Northwestern acknowledges support by the National Science Foundation under grants AST-1814782, AST-1909358 and CAREER grant AST-2047919. WF gratefully acknowledges support by the David and Lucile Packard Foundation.
EB acknowledges support from NSF and NASA grants. PV acknowledges support from NASA grant NNM11AA01A. SB is supported by a Dutch Research Council (NWO) Veni Fellowship (VI.Veni.212.058). Support for this work was provided by the National Aeronautics and Space Administration through \Chandra\ Award Numbers GO1-22059X and DD1-22132X issued by the Chandra X-ray Center, which is operated by the Smithsonian Astrophysical Observatory for and on behalf of the National Aeronautics Space Administration under contract NAS8-03060. The scientific results reported in this article are based in part on observations made by the \Chandra\ X-ray Observatory. This work is based on observations obtained with XMM-Newton, an ESA science mission with instruments and contributions directly funded by ESA Member States and NASA. This work made use of data supplied by the UK Swift Science Data Centre at the University of Leicester. This research is based on observations made with the NASA/ESA Hubble Space Telescope obtained from the Space Telescope Science Institute, which is operated by the Association of Universities for Research in Astronomy, Inc., under NASA contract NAS 5–26555. These observations are associated with program 16303.
This paper makes use of the following ALMA data: ADS/JAO.ALMA\#2019.1.00863.T. ALMA is a partnership of ESO (representing its member states), NSF (USA) and NINS (Japan), together with NRC (Canada), MOST and ASIAA (Taiwan), and KASI (Republic of Korea), in cooperation with the Republic of Chile. The Joint ALMA Observatory is operated by ESO, AUI/NRAO and NAOJ. The Australia Telescope Compact Array is part of the Australia Telescope National Facility which is funded by the Australian Government for operation as a National Facility managed by CSIRO. We acknowledge the Gomeroi people as the traditional owners of the Observatory site. 

\appendix
\restartappendixnumbering

\section{The BAT/GUANO localization}
\label{text:nitrates} 
The highest likelihood position of \grb\ as determined by the NITRATES analysis is close to the edge of the BAT coded field of view with only 3.9\% of the detector plane coded, which precludes localization of the burst via the traditional coded aperture imaging techniques. To illustrate this, we generate BAT sky images with the event data from GUANO, which reveals a source with SNR 3.58 at the best fit position from NITRATES. However, performing traditional image-domain analysis on this BAT sky image, we find that this source is only the 172nd (!) most-likely position for the burst, and thus the event is entirely hidden in the noise in the image domain. 

From NITRATES, the difference in log-likelihood between this best-fit position and other positions in the BAT FOV is $\Delta$LLHPeak=6.7, and between this best-fit position and the best fit out-of-FOV position is $\Delta$LLHOut=7.2. These measure the statistical preference for the specific arcminute-scale position derived by NITRATES compared to other possible positions on the sky, and the confidence that the burst originated from a position within the BAT coded FOV, respectively. Both of these values are on the extreme lower boundary for confident locations that can be derived from BAT data, and thus the position was initially reported as a `candidate' localization \citep{gcn31049}. 

Of all events ever successfully localized with BAT and subsequently confirmed via afterglow discovery to date, \grb\ is the weakest in the image domain. It is also the first short burst discovered and localized by GUANO with a confirmed afterglow. 
This highlights the power of the NITRATES technique in localizing weak (in particular, short) GRBs, which would otherwise be impossible to follow up and characterise, with traditional imaging-based $\gamma$-ray techniques alone. 

\section{Classification model parameters}
\label{text:prob} 
The density of the points in the duration-hardness plane is described by two, two dimensional log-normal distributions. One component has the following form:
\begin{equation}
    f(X)= \frac{\exp\left(-\frac{1}{2}(X-\bar{X})^\top V^{-1} (X-\bar{X})\right)}{\sqrt{(2\pi)^2 \det(V)}}
\end{equation}
where $X$ is a vector composed of ($\log_{10} T_{90}, \log_{10}{\rm HR}$), $V$ is the matrix of variances, $\bar{X}$ is a vector containing the means. To calculate the probability of a GRB being short, we have 
\begin{equation}
    P_{\rm short}(X)=\frac{w_{\rm short}f_{\rm short}(X) }{w_{\rm short}f_{\rm short}(X) + w_{\rm long}f_{\rm long}(X)}
\end{equation}
where $w$ parameters indicate the weight of the two components ($w_{\rm short}=1-w_{\rm long}$). For the component describing the short class in Figure \ref{fig:gamma-ray}, and also in \citet{rfv+21}, we have 
$w_{\rm short}=0.2094$,  $
\bar{X}_{\rm short} = (-0.0256, 0.2018)$ and
\begin{equation}
V_{\rm short}= 
\begin{pmatrix}
0.2779& -0.1037\\
-0.1037 & 0.1354
\end{pmatrix}, 
\end{equation}
while for the long population:
$w_{\rm long}=0.7906$, $
\bar{X}_{\rm long} = (1.4630, -0.1944)$ and
\begin{equation}
V_{\rm long}= 
\begin{pmatrix}
0.2058& -0.01187\\
-0.0119&  0.0511
\end{pmatrix} 
\end{equation}.

\begin{deluxetable}{ccc}
 \centering
 \tabletypesize{\footnotesize}
 \tablecolumns{3}
 \tablecaption{X-ray observations of GRB~211106A}
 \tablehead{   
   \colhead{Time} &
   \colhead{Count\,Rate} &
   \colhead{Unabsorbed Flux}\\[-0.1in]
   \colhead{(days)} &
   \colhead{($10^{-3}$\,counts\,s$^{-1}$)} &
   \colhead{($10^{-14}$\,erg\,s$^{-1}$\,cm$^{-2}$)}
   }
 \startdata 
\multicolumn{3}{c}{\Swift/XRT-PC} \\
\hline
$0.5$  & $12\pm3$ &   $71_{-19}^{+22}$   \\
$0.6$  & $7\pm2$ &   $49_{-12}^{+14}$   \\
$2.7$  & $1.8\pm0.4$ &   $11\pm3$   \\
$27$   & $<5$ &   $<29$   \\
\hline
\multicolumn{3}{c}{\Chandra/ACIS-S3} \\
\hline
$10.5$  & $1.7\pm0.4$ &   $3.7_{-0.6}^{+0.7}$   \\
$59.8$  & $<0.3$  &   $<0.6$   \\
\hline
\multicolumn{3}{c}{\XMM/EPIC} \\
\hline
$14.9$  & $2.6\pm0.7$, $1.6\pm0.4$, $1.1\pm0.4$ &   $2.2\pm0.3$   \\
$33.0$  & $1.0\pm0.4$, $0.3\pm0.2$ &   $1.0\pm0.2$   \\
\hline
\hline
\multicolumn{3}{c}{Best-fit Spectral Parameters} \\
\hline
$z$     &       $\Gamma_{X}$        &       $N_{\rm H, int}$\\
        &                           &       (10$^{21}$\,cm$^{-2}$)\\
\hline
0.5     &   $1.9\pm0.3$  &       $6.3^{+3.7}_{-3.2}$ \\
1       &   $1.9\pm0.3$  &       $13^{+8}_{-7}$ \\
\enddata
\tablecomments{Time is log-centered. \textit{XMM-Newton} count rates are listed per detected in order: pn, MOS1 and MOS2 (first epoch) and pn, MOS1 (second epoch). Fluxes are reported in the 0.3--10\,keV band (observer frame).}
\label{tab:data:xray}
\end{deluxetable}

\begin{deluxetable}{ccccc}
\label{tab:data:radio}
 \tabletypesize{\footnotesize}
 \tablecolumns{5}
 \tablecaption{Radio \& mm observations of GRB~211106A}
 \tablehead{   
   \colhead{Telescope} &
   \colhead{Frequency} &
   \colhead{Time} &
   \colhead{Flux density} &
   \colhead{Uncertainty}\\[-0.1in]
   \colhead{} &
   \colhead{(GHz)} &
   \colhead{(days)} &
   \colhead{($\mu$Jy)} &
   \colhead{($\mu$Jy)}
   }
 \startdata 
ATCA &$5.5$   & $14.18$       & $109$    & $11$ \\
ATCA &$5.5$   & $20.34$       & $139$    & $15$ \\
ATCA &$5.5$   & $27.27$       & $157$    & $32$ \\
ATCA &$5.5$   & $42.05$       & $149$    & $11$ \\
ATCA &$5.5$   & $62.60$       & $121$    & $13$ \\
ATCA &$5.5$   & $117.04$      &  $27$    & $15$ \\
ATCA &$9.0$   & $14.18$       & $130$    & $11$ \\
ATCA &$9.0$   & $20.34$       & $192$    & $13$ \\
ATCA &$9.0$   & $27.27$       & $84$     & $27$ \\
ATCA &$9.0$   & $42.05$       & $106$    & $11$ \\
ATCA &$9.0$   & $62.60$       &  $66$    & $13$ \\
ATCA &$9.0$   & $117.04$      &  $17$    & $12$ \\
ATCA & $18.0$  & $27.25$      & $144$    & $28$ \\
ATCA & $18.0$  & $41.21$      & $<123$   & $41$ \\
ATCA & $18.0$  & $62.72$      & $158$    & $41$ \\
ATCA & $18.0$  & $115.88$     & $<81$    & $27$ \\
ATCA & $34.0$   & $20.20$     & $<372$   & $124$ \\
ATCA & $34.0$   & $27.16$     & $<216$   & $72$ \\
ATCA & $34.0$   & $41.20$     & $<138$   & $46$ \\
ALMA &$97.5$ &$12.89$  &$148$   &$11$ \\
ALMA &$97.5$ &$19.72$  &$141$   &$11$ \\
ALMA &$97.5$ &$27.78$  &$103$   &$12$ \\
ALMA &$97.5$ &$42.70$  &$57$    &$14$ \\
ALMA &$97.5$ &$62.55$  &$<39.6$  &$13.2$
\enddata
\tablecomments{We report mean time post-burst in all cases, including where observations span multiple, adjacent days.}
\end{deluxetable}

\begin{deluxetable}{lllccccc}
\label{tab:data:nir}
 \tabletypesize{\footnotesize}
 \tablecolumns{9}
 \tablecaption{HST NIR observations of GRB~211106A}
 \tablehead{   
   \colhead{Time} &
   \colhead{Instrument} &
   \colhead{Object} &
   \colhead{Band} &
   \colhead{Magnitude} &
   \colhead{Uncertainty} \\[-0.1in]
   \colhead{(days)} &
   \omit &
   \omit &
   \omit &
   \omit &
   \omit
   }
 \startdata
 19.05 & ACS     & AG     & F814W & $>26.00$ & \ldots \\
 19.18 & WFC3/IR & AG & F110W & $>27.01$ & \ldots \\
 25.26 & WFC3/IR & AG & F110W & $>27.01$ & \ldots \\
 Stack & ACS     & H & F814W & 25.791   & 0.069 \\
 Stack & WFC3    & H & F110W & 25.709   & 0.016 
\enddata
\tablecomments{Limits on the afterglow (AG) flux are computed by forced photometry on residual images obtained after subtracting the final epoch (at 48.15 days) from the given epoch. We report photometry of object H identified in the stacks (Fig.~\ref{fig:HSTpanels}) in the last two rows.}
\end{deluxetable}

\section{Refined X-ray astrometry}
\label{text:astrometry}
We derive a refined X-ray afterglow position by registering the \Chandra\ and \HST\ images on a common reference frame. Since there are no sources in common between the two, we proceed via a Legacy Survey image of the field, which we tie to the HST reference frame using 15 sources ($\sigma_{\rm tie,Legacy-HST}=0.03\arcsec$). We tie the \Chandra\ image to Legacy using two common sources ($\sigma_{\rm tie,Legacy,\it Chandra}=0.15\arcsec$). The \Chandra\ position in the \HST\ frame is 
RA = 22h\,54m\,20.518s, Dec = $-53$d\,13\arcmin\,50.590\arcsec, uncertainty $0.18\arcsec$, including the combined uncertainty in the astrometric tie and the centroid uncertainty from \Chandra. This is the circle labeled ``CXO'' plotted in Fig.~\ref{fig:HSTpanels}. 

\section{Host Galaxy SED Fits}
\label{text:cigale}
\begin{figure*}
 \centering
 \includegraphics[width=\textwidth]{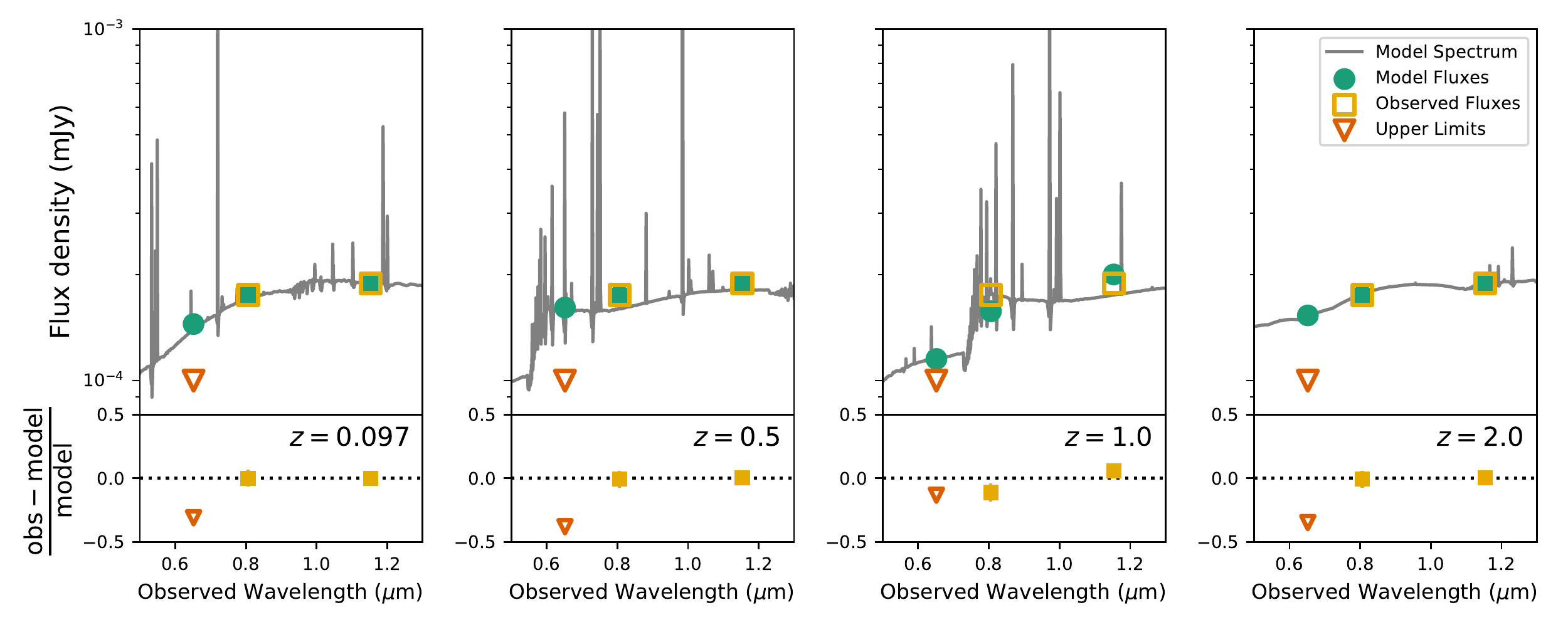}
 \caption{SEDs of the host galaxy of \grb\ in $R$-band (triangle), $F814W$ and $F110W$ (squares), together with SED models from CIGALE at three redshifts (grey lines) with the model fluxes as green circles. The red $R-F814W$ color suggests a redshift, $z\approx1$.}
\label{fig:cigale}
\end{figure*}

We fit the SED of the host galaxy (H) of \grb\ with CIGALE \citep{nbg+09} following \cite{hps+20} at four different redshifts ($z=0.097$, $z=0.5$, $z=1$, and $z=2$) and present the best-fit models, together with the corresponding residuals in Figure~\ref{fig:cigale}. The $z=0.097$ model severely over-predicts the $R$-band upper limit, ruling out a redshift of $z=0.097$ for the host galaxy. The red $R-F814W$ color requires a break between the two bands, which is most easily explained as a 4000\AA\ break, suggesting a redshift of $z\approx0.7$--1.4 (Section~\ref{text:host}). The available photometry is too sparse to further constrain the host galaxy properties. Further photometric or spectroscopic observations (e.g. with JWST) could help constrain important parameters such as the true redshift and the host galaxy extinction.

\section{Model without IC/KN corrections}
\label{text:noKNmodel}
Here we briefly discuss the $z=1$ parameter set without IC/KN corrections. The parameters of the corresponding highest-likelihood model are, $\EKiso\approx3.9\times10^{51}$~erg, $\dens\approx8.0\times10^{-3}~\pcc$, $\epse\approx0.24$, $\epsb\approx0.75$, $\tjet\approx32$~days,  $p\approx2.15$, and $A_{\rm V}\gtrsim3.2$~mag. 
The break frequencies at $\approx1$~day are $\numax\approx3.9\times10^{12}$~Hz and $\nuc\approx2.0\times10^{15}$~Hz. Like in the model incorporating IC/KN effects, $\nua$ is below the radio band and is not constrained. The flux density at $\numax$ is $\fnumax\approx0.14$~mJy. This model yields a slightly worse fit to the X-ray and ALMA light curves, but is otherwise similar to the $z=1$ model with IC/KN effects presented above. The values of $\tjet$ and $\thetajet$ and the limits on $A_{\rm V}$ in this model are similar to those inferred when including IC/KN effects. The value of $\epse/\epsb\approx0.3$ is very different from the value of $\approx10^5$ for the IC/KN model, which is expected, as this ratio is proportional to the Compton $Y$ parameter and IC/KN effects are only important for $Y\gtrsim1$. We note that previous studies have been unable to constrain these microphysical parameters individually in almost all cases due to a paucity of data, and have usually assumed fiducial values (e.g.~ $\epse=0.1$ and $\epsb=0.1$ or $0.01$) for them. 
The best fit and median MCMC values for $\EKiso$ and $\dens$ for this model are comparable to their median values for SGRBs, also derived without including IC/KN effects \citep{fbmz15}. However, the beaming-corrected kinetic energy, $\EK\approx1.6\times10^{50}$, remains at the extreme high end of the distribution for $\EK$ (Section~\ref{text:luminosity}).

\section{Impact of the Klein-Nishina Correction}
\label{text:KN}
In the highest likelihood $z=1.0$ model, the value of the electron index is constrained to $p=2.47\pm0.05$. This is steeper than that derived by applying standard closure relations to the X-ray light curve and the difference can be explained by IC cooling. For the highest likelihood parameters, we find $Y_{\rm c}\approx280$ with $\nuc<\nux$; however, this value decreases with time, resulting in non-standard light curve evolution, since $\nuc\propto(1+Y_{\rm c})^{-2}$. For these parameters, IC cooling is weakly KN suppressed, and the spectral ordering at $\approx1$~day is $\numax<\nuchat\lesssim\nuc<\nux$, where $\nuchat$ is the KN break corresponding to electrons unable to cool efficiently by IC emission while radiating above $\nuc$ \citep{nas09}. The expected spectral index in this regime is $\beta=3(1-p)/4\approx-1.1$, consistent with the observed X-ray spectral index, $\beta_{\rm X}=-0.92\pm0.30$. The expected light curve\footnote{The first term arises from the different spectral index above $\nuc$ and the second from the evolution of $Y_{\rm c}$ with time.} in this regime is $\alpha=7(1-p)/8+(p-2)/2\approx-1.0$ \citep{nas09,lab+18}, which is consistent with the observed value of $\alpha_{\rm X}=-0.97\pm0.03$. We note that a similar slower evolution of the X-ray light curve in GRB~161219B was previously explained as arising from the same spectral regime \citep{lab+18}, although here we also incorporate the effects of an evolving $Y_{\rm c}(t)$. 

\bibliographystyle{apj}
\bibliography{main_rev2_astroph}

\end{document}